\documentclass[12pt,preprint]{aastex}
\usepackage{emulateapj5,graphicx,times}

\slugcomment{AJ accepted}
\shorttitle{Extended emission line regions in Coma}
\shortauthors{Yagi et al.}

\def\Ha{H$\alpha $}

\def\NB{NB}

\begin{document}

\title{A dozen new galaxies caught in the act:
Gas stripping and extended emission line regions in the Coma cluster
\footnotemark[1]}

\footnotetext[1]{Based on data collected at the Subaru Telescope, 
which is operated by the National Astronomical Observatory of Japan.}

\author{
Masafumi Yagi\altaffilmark{2,3},
Michitoshi Yoshida\altaffilmark{4},
Yutaka Komiyama\altaffilmark{3},
Nobunari Kashikawa\altaffilmark{3},
Hisanori Furusawa\altaffilmark{5},
Sadanori Okamura\altaffilmark{6,7},
Alister W. Graham\altaffilmark{8},
Neal A. Miller\altaffilmark{9},
David Carter\altaffilmark{10},
Bahram Mobasher\altaffilmark{11},
Shardha Jogee\altaffilmark{12}
}

\altaffiltext{2}{email:YAGI.Masafumi@nao.ac.jp}
\altaffiltext{3}{
Optical and Infrared Astronomy Division,
National Astronomical Observatory of Japan,
2-21-1, Osawa, Mitaka, Tokyo, 181-8588, Japan}
\altaffiltext{4}{
Hiroshima Astrophysical Science Center, Hiroshima University,
1-3-1, Kagamiyama, Higashi-Hiroshima, Hiroshima, 739-8526, Japan
}
\altaffiltext{5}{
Astronomy Data Center,
National Astronomical Observatory of Japan,
2-21-1, Osawa, Mitaka, Tokyo, 181-8588, Japan}
\altaffiltext{6}{Department of Astronomy, School of Science,
University of Tokyo, 7-3-1 Hongo, Bunkyo-ku, Tokyo 113-0033, Japan}
\altaffiltext{7}{Research Center for the Early Universe, 
University of Tokyo, 7-3-1 Hongo, Bunkyo-ku, Tokyo 113-0033, Japan}
\altaffiltext{8}{Centre for Astrophysics and Supercomputing,
Swinburne University of Technology, Hawthorn, Victoria 3122, 
Australia
}
\altaffiltext{9}{
Department of Astronomy, University of Maryland,
College Park, MD, 20742-2421, USA}
\altaffiltext{10}{
Astronomical Research Institute, Liverpool John Moores University,
Twelve Quays House, Egerton Warf, Birkenhead CH41 1LD, UK
}
\altaffiltext{11}{
Department of Physics and Astronomy, University of California, 
Riverside, CA 92521, USA
}
\altaffiltext{12}{
Department of Astronomy, University of Texas at Austin
1 University Station C1400, Austin, TX 78712, USA
}


\begin{abstract}

We present images of extended \Ha\ clouds associated with 
14 member galaxies in the Coma cluster
obtained from deep narrow band imaging observations with the 
Suprime-Cam at the Subaru Telescope. The parent galaxies of the extended 
\Ha\ clouds are distributed farther than 0.2 Mpc from 
the peak of X-ray emission of the cluster.
Most of the galaxies are bluer than $g-r \approx 0.5$ 
and they account for 57\%\ of the blue ($g-r<0.5$) bright 
($r<17.8$ mag) galaxies in the central region of the Coma cluster. 
They reside near the red- and blue-shifted edges
of the radial velocity distribution of Coma cluster
member galaxies.
Our findings suggest that the most of the parent galaxies were 
recently captured by the Coma cluster potential and are now 
infalling toward the cluster center with their disk gas 
being stripped off and producing the observed \Ha\ clouds.
\end{abstract}



\section{Introduction}

Clusters of galaxies are ideal laboratories for investigating
environmental effects on galaxy evolution. It is well known
that galaxies residing in the central region of clusters are
generally deficient in gas
\citep[e.g.,][]{Giovanelli1985, Gavazzi1987, Solanes2001,
Boselli2006, BravoAlfaro2009}. 
This indicates that some gas
removal mechanisms work quite efficiently in the cluster center. 
Recently, \citet{vanderWel2010} studied the shape and magnitude of 
quiescent galaxies in SDSS and found that galaxies in dense 
environments include gas-removed spiral galaxies, 
adding to the evidence for rapid gas removal.
A number of possible physical processes have been
proposed to explain efficient gas removal in clusters: 
ram pressure \citep{Gunn1972} or
viscous stripping \citep{Nulsen1982}, 
tidal stripping by galaxy-galaxy interactions
\citep{Toomre1972}, 
or the cluster potential
\citep{Byrd1990}, 
evaporation \citep{Cowie1977},
and to a lesser degree,
galaxy harassment \citep{Moore1996}.

One critical clue for understanding 
the gas removal is detection of the stripped gas.
Recent deep \Ha\ imaging observations revealed that 
some of the galaxies in clusters are accompanied by
\Ha\  clouds.
In the nearby cluster Abell 1367,
\citet{Gavazzi2001} found extended ionized gas regions
associated with starburst irregular galaxies.
The sizes of the regions are 75$\times$8 and 50$\times$8 kpc.
\citet{Yoshida2002} discovered another
extended emission line region ($\sim$35 kpc) in the Virgo cluster,
which was found to be a part of 110$\times$25 kpc
of HI gas \citep{Oosterloo2005}.
\citet{Yagi2007} reported a narrow and long (60$\times$2 kpc)
\Ha\ emitting region associated with a galaxy of the Coma cluster.
\citet{Sun2007} found
a $\sim$ 40 kpc \Ha\ tail of a galaxy in Abell 3627.
The galaxy also had a $\sim$ 70 kpc X-ray tail.
\citet{Sun2010} investigated the galaxy to show
two of $\sim$ 80 kpc tails, and also reported one more galaxy 
with $>$20 kpc \Ha\ tail in the cluster.
\citet{Kenney2008} detected 120 kpc \Ha\ filaments 
from NGC 4438, a spiral galaxy in the Virgo cluster.
\citet{Yoshida2008} reported a galaxy with detached 
\Ha\ clouds with young blue stars in the Coma cluster,
which resembles two galaxies in z$\sim$0.2 clusters 
reported by \citet{Cortese2007}.
These \Ha\ cloud findings imply that 
gas removal mechanisms to form such 
extended emission line regions are
common in clusters of galaxies.
However, it is not yet clear 
how common this phenomenon is.

The Coma cluster is the nearest rich cluster of galaxies,
and is the best target for investigating the \Ha\ clouds 
in the cluster environment with a deep and spatially 
resolved observation.
We have performed a deep \Ha\ narrow band imaging survey
of the Coma Cluster using the Suprime-Cam at the Subaru Telescope
and have already reported some of the prominent clouds detected 
\citep{Yagi2007,Yoshida2008}.
Here, we report significant, extended \Ha\ clouds
around further 12 galaxies, increasing the sample to 
a total of 14.
This is the first paper in a series 
providing a statistical discussion of extended emission line regions
in the central region of the Coma cluster.
The nature of the clouds and physical interpretations
will be presented in subsequent papers.

We assume that the distance modulus 
of the Coma cluster is $(m-M)_0=35.05$ \citep{Kavelaars2000} 
and ($h_0$,$\Omega_M$,$\Omega_\lambda$)=(0.71,0.27,0.73) 
\citep{Larson2010}.
Under these assumptions, 
the angular diameter distance is 97.47 Mpc, and 
1 arcsec corresponds to 0.473 kpc.

\section{Observation and Data Reduction}

\begin{figure}[h]
\includegraphics[scale=0.55,bb=20 17 592 779]{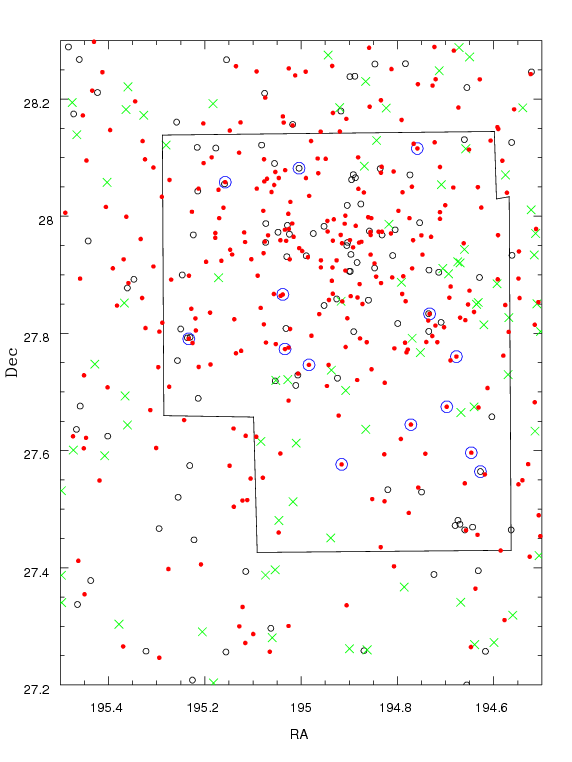}
\caption{
Coma cluster area.
The solid line represents the region observed in the present study.
The symbols are SDSS galaxies of $r<18$.
The red filled circles have a spectroscopic redshift 
between 0.015 and 0.035, the green crosses have a 
redshift out of the range,
and the black open circles have no SDSS spectroscopic information.
The blue large open circles show the parent galaxies of 
the ionized clouds.
Two of the parents did not have SDSS redshift, and 
were confirmed to be Coma members by spectroscopy of Hectospec.
}
\label{fig:FC}
\end{figure}

We observed a field around the Coma cluster center
($\alpha,\delta$)(J2000.)=(12:59:26,+27:44:16)
with Suprime-Cam \citep{Miyazaki2002}
at the Subaru Telescope in 2006-2009.
The field position and orientation are shown in Figure \ref{fig:FC},
and the observations are summarized in Table \ref{tab:obs}.
The observed region was within 1 Mpc from the center,
which is less than half of the virial radius of 
the Coma cluster \citep[2.8 Mpc;][]{Okabe2010}.
We used three broadband filters (B, R, i), and a 
narrowband filter (N-A-L671, hereafter \NB).
The NB-filter was designed for observing \Ha\ emitting
objects in the Coma cluster at z=0.0225,
and has bell-shaped transmission with a central wavelength of
6712 \AA\ and FWHM of 120 \AA\, \citep{Yagi2007,Yoshida2008}.

The imaging data were reduced in a standard manner.
Overscan subtraction, flat-fielding, distortion correction,
background subtraction and mosaicing were performed
using another implementation \citep[nekosoft;][]{Yagi2002}
of the standard Suprime-Cam reduction software.
Satellite trails and contaminated pixels with saturated photons 
were masked manually before mosaicing.
The flux calibration was performed using 
the seventh Data Release \citep[hereafter DR7]{DR7}
of the Sloan Digital Sky Survey (SDSS)
photometric catalog.
The details are described in Appendix \ref{sec:photocal}.

The limiting surface brightness and 
PSF sizes of the final B, R, i, and \NB\ 
images are listed in Table \ref{tab:obs}.
The limiting surface brightness was estimated by
photometry of randomly sampled apertures 2 arcsec in diameter.
We used the median of an absolute deviation (MAD) multiplied 
by a factor $f=1.482602$ as a robust estimator of the rms.
The factor was calculated by solving erf(1/f)=3/4
\[
\int_{-\infty}^{1/f} \frac{1}{\sqrt{2\pi}}e^{-\frac{x^2}{2}} dx=3/4,
\]
and $f\times$MAD equals to the rms,
when the error followed a Gaussian distribution.

\section{Detection of \Ha\ Clouds}

\subsection{\Ha\ image}

\begin{figure}[h]
\includegraphics[angle=-90,scale=0.3,bb=22 16 596 784]{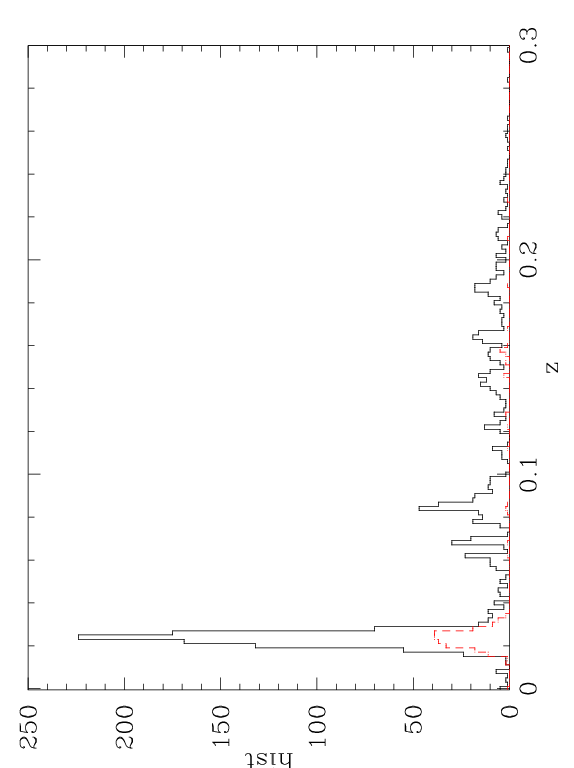}
\includegraphics[angle=-90,scale=0.3,bb=22 16 596 784]{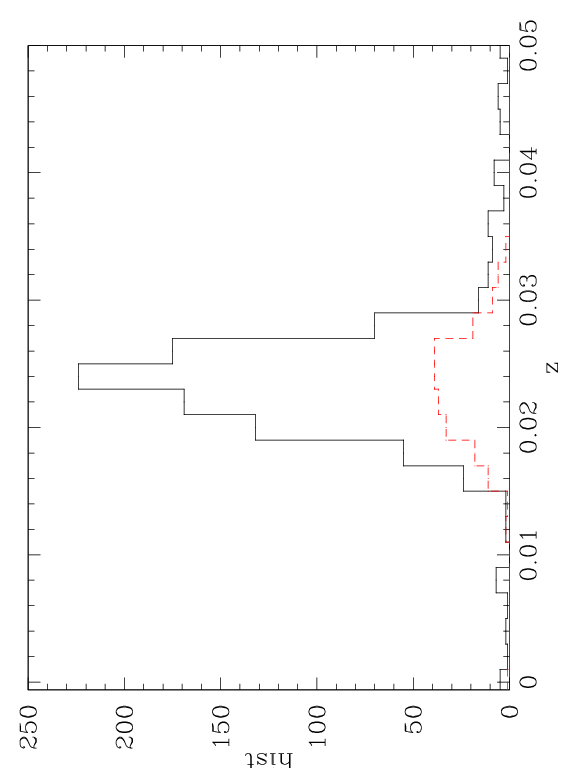}
\caption{
Redshift distribution of SDSS galaxies brighter than $r=17.8$ mag
that were located in our field (red broken line histogram)
and within a 2-degree radius from the cluster center 
(black solid line histogram).
The right figure is a zoom around z=0.025.
}
\label{fig:Nz}
\end{figure}

Figure \ref{fig:Nz} shows the redshift distribution of 
galaxies brighter than $r=17.8$ mag within a radius of 
2 degrees from the cluster center,
and those in our observed field.
Redshifts were taken from  SDSS DR7
spectroscopic catalog.
Based on Fig. \ref{fig:Nz},
we defined the galaxies with $0.015<$z$<0.035$ as
Coma member galaxies.

\begin{figure}[ht]
\includegraphics[angle=-90,scale=0.3,bb=22 16 596 784]{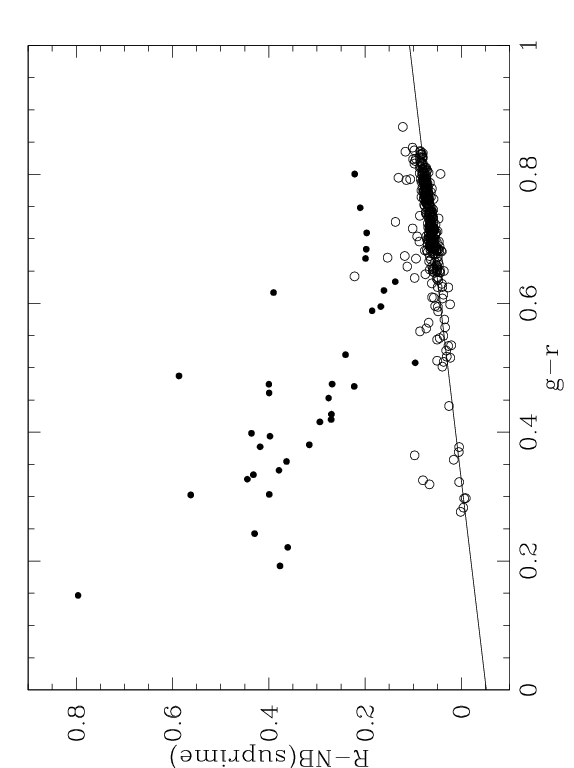}
\includegraphics[angle=-90,scale=0.3,bb=22 16 596 784]{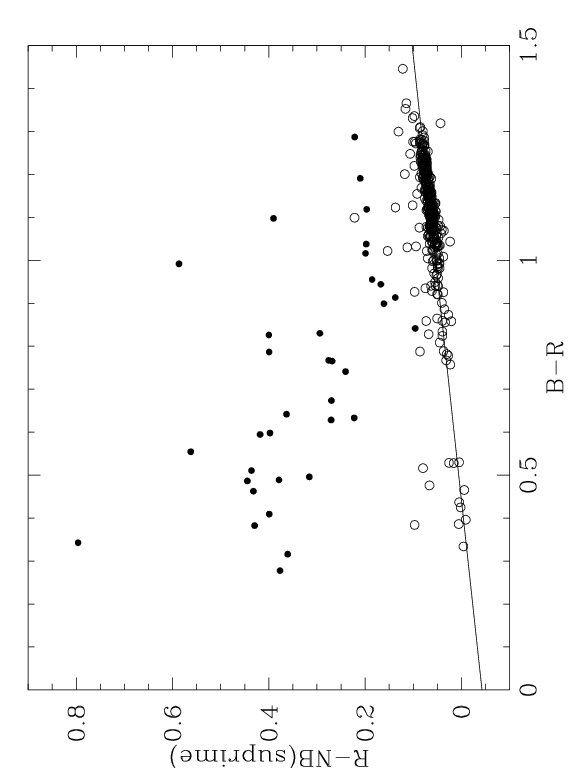}
\caption{
$R-\NB$\ color of cluster member galaxies 
estimated from the SDSS spectrum 
as a function of SDSS $g-r$ color using fiberMag (left)
and Suprime-Cam $B-R$ color estimated from the SDSS spectrum of 
each galaxy (right).
The dots represents those with EW(\Ha)$>$10\AA\ and 
the open circles represent other galaxies.
The solid line is the regression line for EW(\Ha)$<$1\AA\ galaxies.
}
\label{fig:member_gr_RHa}
\end{figure}

To detect \Ha\ clouds,
we subtracted the R-band image from the \NB\ image.
Here, we refer to the R-band subtracted \NB-band image as 
the \Ha\ image. 
First, the typical $R-\NB$ color of Coma members was estimated
by multiplying the filter response to the SDSS spectra
of galaxies with $0.015<$z$<0.035$.
There are 395 galaxies in the redshift range 
with SDSS spectra in our observed field.
The results are shown in Figure \ref{fig:member_gr_RHa},
where $R-\NB$ color is plotted as a function of SDSS color $(g-r)$
and Suprime-Cam color $(B-R)$.
As the $R-\NB$ is calculated from the SDSS spectrum, we 
use fiberMag for $(g-r)$, which corresponds to the color
of the central 3-arcsec aperture.
The large dots represent the galaxies with equivalent width
EW(\Ha)$>$10\AA\ in SDSS.
Figure \ref{fig:member_gr_RHa} shows that the 
$R-\NB$ color of galaxies with EW(\Ha)$<$10\AA\ 
correlates with the $g-r$ and $B-R$ color.
The lines are the regression line for EW(\Ha)$<$1\AA\ galaxies:
\[
R-\NB = 0.15 (g-r)_{{\rm SDSS}} - 0.043
\]
or
\[
R-\NB = 0.097 (B-R) - 0.043.
\]

This relation is useful in the analysis of emission line galaxies.
The main purpose of this study, however, 
was to detect nebulous features,
where neither $g-r$ information nor our $B-R$ color 
are necessarily available with a reliable S/N ratio.
Therefore, we used the median color of these member galaxies,
$R-\NB$=0.065 mag, as the color of a typical Coma passive galaxy,
i.e., a galaxy with no \Ha\ emission. 
We scaled the R-band image so that such galaxies had
values of 0 in the \Ha\ image.
As slight ($<$1 pix, 0.2 arcsec) positional shifts
between the R-band and \NB-band remained in some regions,
sub-pixel alignment was performed by the method of 
\citet{Yagi2010} before subtraction.

We also constructed a three-color image from B, R, and \NB.
The colors of the member galaxies in B, R, i, and \NB\, in 
the Suprime-Cam filter system were calculated from SDSS spectra.
We adopted a scale such that those galaxies with
$B-R$=1.0, $R-\NB$=0.065, and $\NB-i$=0.2
were gray in the three-color image.

In the \Ha\ image, clouds were searched by visual inspection.
Ghosts of bright stars and residuals of 
background subtractions were carefully checked and excluded.
The correlation of $R-\NB$\ and color 
(Figure \ref{fig:member_gr_RHa})
suggested that red galaxies would have residuals in the \Ha\ image,
while blue galaxies would have negative values.
This point was verified using B and i-band data.
If a region was bright in \NB\ and also red in the 
$B-R$ and $R-i$ colors, it was regarded as a residual. 
If the color was blue, however,
it was regarded as a star-forming region, 
and was included in our sample of \Ha\ clouds.

\subsection{List of clouds and parameter measurements}

All of the extended \Ha\ clouds detected here
are shown in Figure \ref{fig:postage2374}-\ref{fig:postage4232} 
and listed in Table \ref{tab:parents}.
We named the clouds after possible associated
galaxies (hereafter, parent galaxies), 
using the galaxy IDs of \citet{Godwin1983}.
The contrast of three-color composites (left images) 
was selected arbitrarily according to the brightness of 
the parent galaxy, while the color balance was fixed.
A pure red color represented \Ha\ emission without any
stellar light, purple represented \Ha\ emission
with some stellar light, and blue color represented
light from young blue stars without any \Ha\ emission.
In Figures \ref{fig:postage3016} and \ref{fig:postage4060},
we can see blue filaments from the parent galaxy in the left image,
while no \Ha\ emission is present in the right image. 
These are the ``fireball'' features reported by
\citet{Yoshida2008}.

The contrast of the \Ha\ image (right images) is the same for 
Figures \ref{fig:postage2374}-\ref{fig:postage4232}.
Therefore, the gray colors could be compared directly
among the figures to measure the apparent \Ha\ brightness.
The green line in the \Ha\ image represents the isophote of 
the \Ha-image at
2.5$\times 10^{-18}$ erg s$^{-1}$ cm$^{-2}$ arcsec$^{-2}$,
assuming that the clouds have the same redshift as the parent.
The red contour represents the isophote of our R-band image,
which was useful for comparing the extension of clouds 
with stellar light.
The R-band isophote was chosen so that the region inside it 
contained the same area as the aperture of the r-band half 
light radius 
(PetroR50) of SDSS, because the bright galaxies in our R-band image 
had saturated pixels and we were not able to measure the radius 
in our data directly. 
In addition, the isophotal treatment was appropriate as 
the parents had distorted shapes.

We performed spectroscopic observations of some of the clouds
using the Faint Object Camera and Spectrograph
\citep[FOCAS,][]{Kashikawa2002} in multi-object spectroscopy (MOS) mode
on 23 Jun 2006, and 19 May 2009 UT.
The case of GMP2910(D100) was reported previously
in \citet{Yagi2007}. Other cloud results 
the clouds will be described in Yoshida et al. (in preparation).
In this study, we used the spectroscopic result
indicating that four clouds had similar redshifts to
their respective parent galaxies 
(GMP2559, GMP2910, GMP2923, and GMP4060).
In Figure \ref{fig:postage2923}, GMP2945, which lies 
just to the west of GMP2923, apparent to be the parent of the clouds,
but GMP2945 had a markedly different redshift (z=0.0207)
from the clouds (z$\sim$ 0.028). 
We identified GMP2923(z=0.0291) as the parent and 
GMP2945 as a chance overlap.
Twelve of fourteen parent galaxies 
had SDSS spectral information, and the other two had
a Hectospec redshift (Marzke, R.O, et al. 2010, in preparation).
There were six galaxies that had both SDSS and Hectospec
redshifts. They were consistent within $\sim$ 30 km $s^{-1}$.


Table \ref{tab:clouds} shows
the extension of the \Ha\ clouds relative to the parent galaxy
and the surface brightness of the clouds
measured in the \Ha\ image.
The \Ha\ total flux was calculated by assuming that
the cloud had the same redshift as the parent galaxy,
and its mix of [NII]6548,6583 was similar to that of the GMP2910 
clouds \citep{Yagi2007}. We adopted a value of [NII]6583/\Ha=0.5. 
Other emission lines, such as [OI]6300 and [SII]6716,6731,
were out of the passband in the redshift of $0.015<$z$<0.035$ 
and were therefore negligible. 
As the wavelength dispersion of the emission lines 
of the GMP2910 cloud in the FOCAS data was $\sim$3.8\AA\, 
\citep{Yagi2007},
we adopted a value for the model to compute the \Ha\ flux.
As it includes broadening by an instrumental profile, 
the value would be overestimated for \Ha\ imaging data.
However the effect of the dispersion was found to be negligible.
The dispersion of 3.8\AA\ corresponded to 
velocity dispersion of $\sim 170$ km sec$^{-1}$,
but even if we assume a dispersion 
as 10 km sec$^{-1}$, the change in \Ha\ magnitude 
is less than 0.01 mag.
We defined the
\Ha\ magnitude (mag(\Ha)) as the continuum subtracted \NB\ magnitude.
\[
mag(H\alpha)_{AB}=-2.5\log\left[\frac{\int R_{\NB} f_\nu d\nu - k \int R_{R} f_\nu d\nu}
{\int R_{\NB} d\nu } \right]- 48.6,
\]
where $R_{\NB}$ and $R_{R}$ are the transmission of the filters,
$f_\nu$ is the flux density, and $k$ is the correction factor 
so that
\[
\int R_{\NB} f_\nu d\nu = k \int R_{R} f_\nu d\nu
\]
for objects with median color without \Ha\ emission, 
$R-\NB=0.065$.
We calculated the mag(\Ha) from the flux in the $\NB-R$ image 
with a zero-point \NB-band magnitude.
It should be noted that the \Ha\ magnitude is not an $\NB-R$ color
but the AB magnitude measured in the \Ha\ image.
That is, the $\NB-R$ color corresponds to EW(\Ha), 
while \Ha\ magnitude corresponds to L(\Ha).

The measurement of flux was performed in the fixed isophote
of a smoothed image. The \Ha\ image was smoothed with a Gaussian kernel
of $\sigma$=2.5 pixel, so that the noise of each pixel was suppressed.
The isophotal threshold for the analysis 
was set to 2.5$\times 10^{-18}$ erg s$^{-1}$ cm$^{-2}$ arcsec$^{-2}$ 
at the redshift of the parent galaxy.
As the transmission of the \NB\ filter has a bell-shaped curve,
the transmission at \Ha\ changes according to the redshift.
This effect was corrected when the threshold was calculated.
The detected objects were examined with three-color images.
Possible overlapping objects, such as foreground stars,
distant galaxies, or a redder galaxy core were removed manually.
Even after this screening, some of the clouds had
errors due to the stellar color of the parent galaxy.
In addition, some included light from HII regions in the parent galaxy.
The examined detection map was used as a mask, and the total flux and
peak surface brightness were measured in the masked \Ha\ image.
The same mask was applied to the R-band and the mean surface brightness
in the R-band (SB$_{R}$), which is an indicator
of the extent of contamination of the \Ha\ emission 
by continuum light, was measured.
The measured values are summarized in Table \ref{tab:clouds}.

\subsection{Note on GMP4017 clouds}
The \Ha\ clouds of GMP4017 lie along the line of 
the clouds of GMP4060, and are slightly elongated in 
the north-south direction (Figure \ref{fig:GMP4017a}).
This implies that they may not be associated with GMP4017,
but with GMP4060. If the GMP4017 clouds were 
in fact an extension of the GMP4060 clouds,
the length of the GMP4060 cloud would be 144 kpc.
As the redshift of GMP4017(0.0280) is similar to that of 
GMP4060(0.0292), it is difficult to identify 
the physical association from spectroscopy. 

HI observations may uncover the origin of the clouds of GMP4017.
However, no HI emission has been detected around GMP4017 by 
previous deep HI surveys \citep{BravoAlfaro2000,BravoAlfaro2001},
with rms per channel of 0.20-0.40 mJy beam$^{-1}$.
Although the survey covered the entire field of this study, 
there were only three parents in which HI gas was detected
(GMP2374, GMP2559, and GMP3779). 
Much deeper HI observations may be required 
to determine whether GMP4017 or GMP4060 is the true parent
of the \Ha\ clouds near GMP4017.

\section{Results}

\subsection{Distribution of the parents in the cluster}
\label{sec:result:pos} 

Figure \ref{fig:FC2} shows the positions of the parents,
the length and the position angle of the clouds
projected onto the sky.
The redshift relative to the Coma cluster is also indicated 
as the color of the symbols.
No clouds extend toward the center, although
they do not always extend in the anti-center direction either.
In addition, there is no apparent sub-clustering of the parents.

Figure \ref{fig:d_z} shows 
the redshift of galaxies 
as a function of the distance from the cluster center.
The position of ($\alpha,\delta$)=(12:59:42.8,+27:58:14) 
was adopted
as the center of the cluster \citep{White1993}.
This coordinate was used in our previous studies
\citep{Komiyama2002,Carter2002}.
Figure \ref{fig:d_z} shows that 
most of these parent galaxies appear to reside in 
the blue and red edges of the redshift distribution,
suggesting that the parents have large velocities
relative to the Coma cluster.
We used the
Kolmogorov-Smirnov(KS) test to check whether 
the distribution of the peculiar velocity with respect to
the cluster mean velocity was the same
between the parents and the other member galaxies.
The sample was restricted in the magnitude range $r<17.8$ 
and in our observed field to ensure unbiased samples.
From the test, the p-value for the null hypothesis that 
the two distributions are the same was calculated as 0.038.
Therefore the distribution of the  parents' redshift 
if significantly different from that of other member galaxies 
with over 95\% confidence.

The parents are also located in the region farther than 0.2 Mpc
from the cluster center avoiding the cluster core.
The difference in the distribution of distance was also 
examined by the KS test.
The distribution of parent galaxies was different from 
that of other member galaxies (p-value=0.045).

Figure \ref{fig:d_z} shows that only three parents lie 
on the blueshifted side while the rest lie on 
the redshifted side. The eleven redshifted parents could be part
of a group falling into the cluster \citep{Poggianti2004}.
However, we could not reject the possibility that the parents 
are falling into
the cluster randomly from both sides, because the probability 
that either the redshifted or blueshifted group
has less than 4 of 14 members is 5.7\%,
given the assumption of random infall.

\subsection{Color of parents}
\label{sec:result:color} 

As the parent galaxies are bright,
the centers of some of the galaxies were 
saturated in our Subaru images.
As the photometric data for these galaxies were not reliable,
we took the photometric data from SDSS DR7.
Figure \ref{fig:colmag} shows the color magnitude diagrams of 
the galaxies in our observed field.
The filled circles represent the SDSS galaxies 
with 0.015$<$z$<$0.035 and r-band magnitude $<$18.0.

It is striking that more than half of the blue galaxies are 
parents of some \Ha\ clouds. 
The color magnitude relation (CMR) of passive 
member galaxies in the Coma cluster was fitted as
\[
g-r=-0.0312\,r + 1.233.
\]
If we define ``blue galaxies'' as 
those bluer by $>$0.2 mag than the CMR 
the fraction of parents in the blue galaxies is 57\% (8/14)
in the $r<$17.8 mag (M$_{r}<-17.25$) SDSS confirmed members.

In the SDSS DR7 photometric catalog, 
there are nine objects in our observed field 
which are classified as galaxies, have a ``blue'' color, 
are brighter than $r<$17.8 mag, and have no SDSS redshift.
We checked the images of those objects and found that
they consisted of three bright stars, one ghost of a bright object,
two parts of other galaxies, and three galaxies.
Two of the three galaxies, GMP3016 and GMP4232,
are members, which is confirmed by Hectospec spectroscopy,
and had \Ha\ tails as shown in Fig 4.
The remaining one was part of the GMP2863 complex.
GMP2863 consists of at least three components,
which we call GMP2863a, GMP2863b, and GMP2863c;
of these, GMP2863b has a blue color in SDSS.
If GMP2863b is a member, the fraction of parents 
in the blue galaxies is 59\%(10/17);
if it is not, the fraction is 63\%(10/16).

It should be noted that we used an apparent $g-r$ color
in this analysis. As we did not correct for internal dust extinction 
of the parent galaxy in this study, 
their intrinsic color may be bluer.

\subsection{Morphology of the cloud-parent connection}

The distributions of \Ha\ clouds around the parent
(Fig \ref{fig:postage2374}-\ref{fig:postage4232})
were varied.
We classified them into three types based on their appearance:
1) connected \Ha\ clouds with disk star formations;
2) connected \Ha\ clouds without disk star formation;
and 
3) detached \Ha\ clouds.
The results are listed in Table \ref{tab:clouds}.

{\bf \#1 connected \Ha\ clouds with disk star formation:}
GMP2374, GMP3271, and GMP3816 show galaxy-wide star formation
and some regions of \Ha\ emission have flown out of the 
galaxies entirely.
GMP3271 and GMP3816 are blue, while
GMP2374 is as red as early-type galaxies and 
lies on the tip of the color-magnitude relation.
Nevertheless, GMP2374 exhibits a very clear spiral pattern
with ongoing star formation at the arms. 
We suppose that GMP2374 could be a progenitor of a passive 
spiral galaxy \citep[e.g.,][]{Goto2003}.
GMP2559 has negative \Ha\ on the northern side, but 
strong \Ha\ emission on the southern side.
We also classified it as belonging to this type.

{\bf \#2 connected \Ha\ clouds without disk star formation:}
GMP2910, GMP3071, and GMP3779 exhibit core star formation
and extended \Ha\ emission connected to the core.
The extended clouds have a blobby structure.
GMP3896 has a similar appearance to type \#1, but 
with more extended emission in the disk.
Their color and magnitude are similar to 
those of type \#1.

{\bf \#3 detached \Ha\ clouds:}
GMP2923, GMP3016, GMP4060, and GMP4232 have no
apparent \Ha\ emission in the parent galaxy,
only outside the galaxy.
The color of the parent is very blue ($g-r=0.27-0.30$), 
and the magnitude is very faint ($r>$16.8).
The four galaxies are in fact the four least luminous 
galaxies among the parents. 
GMP3016 and GMP4060 show fireball features:
\Ha\ emission following knots of young blue stars \citep{Yoshida2008}.

The clouds of GMP4017 and GMP4156 are also detached.
However, the emission is superimposed on the stellar component 
of the parent galaxy, and the parents are red.
The clouds near the two giant galaxies could be 
different from those near the other four dwarfs.
We classified GMP2923, GMP3016, GMP4060, and GMP4232 as 
\#3 blue dwarfs, 
and GMP4017 and GMP4156 as \#3 red giants.
The central $R-\NB\ $ excess of GMP4156 was 
partly due to the saturation of our R-band image and 
red color of stellar components, although
\Ha\ emission existed in the SDSS spectrum of 3 arcsec aperture. 
The SDSS spectrum showed a high [NII]/\Ha\ ratio 
of $\sim$0.85,
which may indicate a weak active galactic nucleus (AGN).

We noticed that all the \#3 blue dwarfs showed post-starburst 
signatures in their SDSS/Hectrospec spectra.
As the fiber sizes were 3 arcsec (1.4 kpc) and 1.5 arcsec (0.7 kpc)
for SDSS and Hectospec, respectively
their central regions are in a post-starburst phase.
Moreover, none of the \#3 blue dwarfs were detected
in deep radio 1.4GHz observations with the Very Large Array 
\citep[VLA;][]{Miller2009}.
This suggests that they now have little or no star formation 
even in their outer regions.
Meanwhile, 7 of 8 \#1 and \#2 parents were detected in radio;
these are star forming galaxies and not radio-loud AGNs.

\section{Discussion}

\subsection{Possible evolution sequence of the parent galaxies}

The findings presented in Sections 
\ref{sec:result:pos} and \ref{sec:result:color} 
suggest that the parent galaxies were recently
captured by the Coma cluster potential and
are now infalling toward the cluster center, while their disk
gas is stripped off and seen as \Ha\ clouds.

The classification based on apparent morphology may imply 
an evolutionary sequence of gas stripping:
a star-forming disk galaxy (\#1) 
evolves into star-forming + 
dead disk phase (\#2) and then the gas clouds are detached (\#3).
Recent numerical simulations of ram-pressure stripping
\citep{Kronberger2008,Kapferer2009,Tonnesen2010}
also showed the evolution of \Ha\ appearance in $\sim$ 500 Myr.
However, it is puzzling that \#3 phase galaxies consist of 
less luminous blue galaxies and bright red giants.
Moreover, GMP2374, the brightest \#1 galaxy, shows a red color.
Although part of the red color is due to possible 
extinction by internal dust, the spectrum of the whole galaxy 
in the GOLDMINE database \citep{Gavazzi2003} 
indicates that it is dominated by
intermediate and old stellar populations 
as inferred from its large D4000 and small CaHK ratio.
Therefore it is difficult to assume that this galaxy 
will evolve into a post-starburst phase with a blue color.
The parents may consist of several kinds of galaxies on
different evolutionary paths.

As a second parameter of various evolutionary paths, 
we estimated the masses of the parents.
Higher mass galaxies should have
larger potential wells and retain some gas for star
formation, while lower mass galaxies might simply evolve from
type \#1 to \#3 without passing through \#2.

The masses of the parents were estimated
by three different methods and are given in Table 4.
\citet{Bell2003} gives stellar mass-to-light ratios (M/L) as 
a function of color. 
They adopted a "diet" Salpeter initial mass 
function (IMF), which has 30\% lower mass than Salpeter IMF
\citep{Salpeter1955}.
We adopted
\[
\log_{10} (M/L)_{\rm i-band} = 0.006 + 1.114(r-i) - 0.15 
\]
and calculated the stellar mass from the SDSS $r$ and $i$ magnitudes.
We neglected K-correction for both the $i$ magnitude and $r-i$ 
colors, which were smaller than 0.1 mag.
The possible effects of internal dust were also neglected.
We also estimated the masses of the parents from the model SED fitting,
using kcorrect v4.1.4 \citep{Blanton2007} with SDSS and public GALEX 
magnitudes 
and redshifts as in \citet{Miller2009}.
The recommended corrections on the kcorrect web page 
\footnote{http://howdy.physics.nyu.edu/index.php/Kcorrect}
were applied, which included
the correction between the AB and SDSS magnitudes,
and the minimum uncertainty of the SDSS magnitude.
The kcorrect used \citet{Chabrier2003} IMF and Padova 1994 
isochrones \citep[and references therein]{Bruzual2003}.
Yet another mass estimation was available in the MPA-JHU SDSS catalog
\footnote{http://www.mpa-garching.mpg.de/SDSS/DR7/}. 
The DR7\_v5.2 version included 11 parents of this study.
We adopted the ``MEDIAN'' column as the estimated mass.
The IMF by \citet{Kroupa2001} was adopted for the 
catalog\citep{Kauffmann2003}.

Although there were variations among the mass estimations of the three 
different methods, we can see that the type \#3 dwarfs 
had less than $2 \times 10^9$ M$_{\sun}$.
On the other hand, the classification between \#1 and \#2 showed
no clear correlation with mass.
In addition, the lowest mass \#3 parents 
($<2 \times 10^9$ M$_{\sun}$) stopped star formation quickly, 
while most of the \#1 and \#2 parents, the 
masses of which were $10^9-10^{11}$ M$_{\sun}$,
continued star formation.
The correlation of the detachment of the clouds,
the activity of star formation, and the mass 
implied that the quick quench of the star formation 
in the \#3 blue dwarfs was due to shallow potentials 
and/or small amounts of gas.
This result regarding the dependency of the star formation
quench on the mass was similar to those in 
previous studies
\citep{Kauffmann2004,Haines2007,Wolf2009,Mahajan2010},
although this is the first time that galaxies of low 
mass ($<10^9$ M$_{\sun}$) have been investigated in this way.

\subsection{Comparison with galaxies with UV asymmetries}

Recently \citet{Smith2010} investigated UV and \Ha\ images of 
the Coma cluster, and listed the galaxies with 
gaseous stripping events (GSE), 
which were detected as UV asymmetries and had colors with
$NUV-i<4$.  
Their observed field overlapped with our field, and there 
were six galaxies in common 
(GMP2559, GMP2910, GMP3016, GMP3816, GMP4060, and GMP4232).
It is striking that the UV asymmetric galaxies 
in our fields are only the six.
This suggests that 
GSE galaxies have extended \Ha\ clouds with high probability.

The other eight parents were not listed as GSE galaxies,
although all of the parents in this study satisfied the 
selection criteria of \citet{Smith2010} ($i<18$, and $0.01<$z$<0.043$).
As the NUV-optical color is a function of 
the star formation strength 
and the age since the last star formation \citep[e.g.][]{Kaviraj2007},
one possibility is that the non-GSE parents are older.
This can explain the red \#3 galaxies,
as age quickly makes the $g-r$ color redder.
The other five star-forming galaxies and a post-starburst galaxy 
have young stars, and the age cannot explain the 
non-detection of UV asymmetry.
It is possible that the UV asymmetric feature is 
faint because the star formation is/was weaker than other cases.
It is also possible that the UV emitting stars have 
a symmetric distribution, while the \Ha\ is asymmetric.
The last case suggests that the distributions of the
young stars and gas differ.
The sign of the different distribution was shown 
in our previous studies \citep{Yagi2007,Yoshida2008}.
This also suggests that the source of ionization is not always 
the UV photons from young stars.
A detailed discussion of the source of ionization will be 
presented in our next paper (Yoshida et al. in preparation).

If we assume that all galaxies from which gas was removed
experienced an \Ha\ parent phase, 
we can speculate on the evolution of the parents.
With regard to infalling galaxies, \citet{Graham2003} investigated 
two disk dwarf galaxies in the Coma cluster, 
which still show spiral arm patterns.
The galaxies, GMP3292 and GMP3629, are thought to be 
transforming from spiral to passive galaxies.
\citet{Graham2003} also noted that these galaxies have 
large peculiar velocities relative to the Coma center.
However, these two galaxies are, 
much redder than parents in this study 
($r=16.5, g-r=0.68$ for GMP3292 and 
$r=18.3, g-r=0.64$ for GMP3629),
but are of comparable mass 
(4.1$\times 10^9$ and 8$\times 10^8$ M$_{sun}$),
with values estimated in the same way as for the parents.
They might be recently evolved shapes from the parents, 
as they still retain the arm pattern.


\section{Summary}

From the Suprime-Cam observation of the center of the Coma cluster,
we found ionized \Ha\ clouds near 14 galaxies (parents).
All the parents were confirmed to be 
Coma member galaxies by
SDSS and Hectospec spectroscopy.
The parent galaxies were distributed 
in the region farther than 0.2 Mpc 
from the cluster center, and tended to reside on 
the blue and red edges of the redshift
distribution, avoiding the mean cluster redshift. 
The majority of the parent galaxies were blue, and
57\% of the blue galaxies brighter than $r=17.8$ 
contained \Ha\ clouds. 

We classified the connection between the parent and the clouds
into three types:
connected clouds with disk star formation,
connected clouds without disk star formation, and detached clouds.
The parents of detached clouds consisted of two groups;
blue dwarfs and red giants. 
The blue dwarfs were the four least luminous galaxies 
among the 14 parents,
and showed no sign of current star formation but 
appeared to be post-starburst. 
The parents with connected clouds showed no significant differences 
regardless of whether they had disk star formation or not.

A comparison with galaxies with UV asymmetries showed that
all such galaxies are accompanied with \Ha\ clouds. 
This suggests that a galaxy undergoing a 
gaseous stripping event has 
extended \Ha\ emitting clouds with high probability.

\acknowledgments

We thank the anonymous referee for suggestions and comments 
that improved this paper.
We thank Dr. Russel J. Smith for information on
their gaseous stripping event galaxies and the suggestive comments,
Dr. H. Yamanoi for discussion on 
the photometric calibration of the Suprime-Cam data,
and Dr. M. Sun for informing us about their latest results.
This work was based on data collected using the Subaru Telescope,
which is operated by NAOJ. 
We are grateful to the Subaru staff for their help.
This work was supported by KAKENHI 21540247.
S.J. acknowledges support from the National Aeronautics and Space
Administration (NASA) LTSA grant NAG5-13063, NSF grant AST-0607748,
and $HST$ grant G0-10861 from STScI, which is operated by
AURA, Inc., for NASA, under NAS5-26555.
This work made use of 
SAOImage DS9\citep{Joye2003},
the NASA's Astrophysics Data System\footnote{http://ads.nao.ac.jp/},
the NASA/IPAC Extragalactic Database (NED)
\footnote{http://nedwww.ipac.caltech.edu/}, 
the SDSS skyserver\footnote{http://cas.sdss.org/},
and the data analysis system at the Astronomy Data Center of NAOJ
\footnote{http://www.adc.nao.ac.jp/}.


\clearpage
\begin{table}
\begin{tabular}{|c|c|c|c|c|}
\hline
filter &
PSF size & SB$_{\rm lim}$(ABmag arcsec$^{-2}$) &
Date(UT) & exposure \\
\hline
B & 1''.06 & 28.6 & 2006-04-28 &  2$\times$450sec  \\
    &        &      & 2006-05-03 &  3$\times$450sec  \\
    &        &      & 2007-05-13 &  4$\times$600sec  \\
    &        &      & 2009-05-26 &  6$\times$600sec  \\
\hline
R & 0''.75 & 28.1 & 2006-04-28 & 11$\times$300 sec\\
    &        &      & 2006-05-03 &  1$\times$60  sec\\
    &        &      & 2007-05-11 &  1$\times$60  sec\\
    &        &      & 2007-05-13 & 12$\times$300 sec\\
    &        &      & 2007-05-14 &  5$\times$360 sec\\
    &        &      & 2007-05-15 &  1$\times$300 sec\\
    &        &      & 2009-05-26 &  4$\times$300 sec\\
    &        &      & 2009-05-27 &  4$\times$300 sec\\
\hline
i & 1''.10 & 27.5 & 2006-04-28 & 4$\times$240 sec  \\
    &        &      & 2006-05-03 & 1$\times$240 sec + 2$\times$120 sec \\
    &        &      & 2009-05-25 & 6$\times$300 sec  \\ 
\hline
\NB & 0''.76 & 27.5 & 2006-04-28 &  8$\times$1800 sec \\
    &        &      & 2006-05-03 &  3$\times$120  sec \\
    &        &      & 2007-05-12 &  9$\times$1800 sec \\
    &        &      & 2007-05-15 &  1$\times$300  sec \\
    &        &      & 2009-05-26 &  4$\times$900  sec \\
\hline
\end{tabular}
\caption{Summary of mosaiced images and observation log.
SB$_{\rm lim}$ represents 1-$\sigma$ fluctuation of
flux in an aperture of 2 arcsec diameter}
\label{tab:obs}
\end{table}

\clearpage

\begin{table}
\begin{tabular}{|c|c|c|c|c|c|c|c|}
\hline
Name&
RA &
DEC&
$r$(AB)&
$g-r$(AB)& 
z& 
other name\\
(1) & (2) & (3) & (4) & (5) & (6) & (7) \\
\hline
GMP2374& 195.233579 &27.790860& 13.0 & 0.83 &0.0267 &  D82,NGC4911\\
GMP2559& 195.157759 &28.058088& 15.1 & 0.47 &0.0255 &  D169,IC4040\\
GMP2910& 195.038083 &27.866444& 15.5 & 0.46 &0.0177 &  D100,Mrk60\\
GMP2923& 195.033500 &27.773172& 16.9 & 0.28 &0.0291 & -\\
GMP3016& 195.004583 &28.082028& 17.8 & 0.30 &0.0259\tablenotemark{b} & - \\
GMP3071& 194.984000 &27.746472& 16.2 & 0.32 &0.0299 &  D66\\
GMP3271& 194.915873 &27.576482& 15.8 & 0.38 &0.0167 &  D41\\
GMP3779& 194.771958 &27.644361& 14.4 & 0.40 &0.0181 &  D54,Mrk58,RB219\\
GMP3816& 194.758750 &28.115556& 15.2 & 0.42 &0.0314 &  D195,NGC4858\\
GMP3896& 194.733181 &27.833384& 14.2 & 0.67 &0.0252 &  D89,IC3949\\
GMP4017& 194.697506 &27.674734& 13.9 & 0.83 &0.0280 &  D58,NGC4854\\
GMP4060& 194.677458 &27.760514& 16.8 & 0.28 &0.0292 &  RB199\\
GMP4156& 194.646614 &27.596389& 13.5 & 0.66 &0.0257 &  D43,NGC4853\\
GMP4232& 194.627917 &27.564306& 17.7 & 0.27 &0.0243\tablenotemark{b}  &  -\\
\hline
\end{tabular}
\caption{List of parent galaxies\tablenotemark{a}
The data are taken from SDSS and NED.}
\label{tab:parents}.
\tablenotetext{a}{
(1) GMP name,
(2) RA(J2000),
(3) Dec(J2000),
(4) $r$-band model magnitude from SDSS DR7,
(5) $g-r$ color from SDSS,
(6) redshift from SDSS,
and
(7) other name
}
\tablenotetext{b}{Hectospec redshift}

\end{table}

\begin{table}
\begin{tabular}{|c|c|c|c|c|c|c|c|c|}
\hline
Name&
area&
extension&
position angle&
SB(R)&
peak SB(\Ha)&
total mag(\Ha)&
\Ha\ flux &
type
\\
(1) & (2) & (3) & (4) & (5) & (6) & (7) & (8) & (9)\\
\hline
GMP2374& 2350 & 28 & 249 & 21.3 & 18.6 & 15.0 & 2.3$\times10^{-13}$ &1\\
GMP2559& 1962 & 90 & 183 & 22.9 & 19.7 & 15.7 & 1.1$\times10^{-13}$ &1\\
GMP2910& 498  & 60 & 60  & 23.0 & 19.2 & 17.7 & 1.9$\times10^{-14}$ &2\\
GMP2923& 130  & 54 & 254 & 25.9 & 25.8 & 21.6 & 6.6$\times10^{-16}$ &3\\
GMP3016& 21   & 34 & 40  & 26.5 & 25.7 & 22.9 & 1.4$\times10^{-16}$ &3\\
GMP3071& 168  & 35 & 247 & 23.5 & 22.9 & 20.8 & 1.5$\times10^{-15}$ &2\\
GMP3271& 124  & 4.7& 157 & 21.2 & 20.8 & 18.1 & 1.4$\times10^{-14}$ &1\\
GMP3779& 436  & 47 & 245 & 21.7 & 19.1 & 16.9 & 3.7$\times10^{-14}$ &2\\
GMP3816& 584  & 21 & 333 & 22.4 & 19.5 & 16.4 & 1.0$\times10^{-13}$ &1\\
GMP3896& 585  & 34 & 234 & 21.2 & 20.0 & 16.8 & 3.9$\times10^{-14}$ &2\\
GMP4017& 38   & 20 & 288 & 25.3 & 25.6 & 22.5 & 2.4$\times10^{-16}$ &3(red)\\
GMP4060& 116  & 81 & 165 & 26.2 & 23.4 & 20.9 & 1.2$\times10^{-15}$ &3\\
GMP4156& 118  & 15 & 26  & 22.4 & 25.8 & 21.3 & 6.5$\times10^{-16}$ &3(red)\\
GMP4232& 92   & 22 & 151 & 25.0 & 22.8 & 20.7 & 1.0$\times10^{-15}$ &3\\
\hline
\end{tabular}
\caption{Properties of \Ha\ clouds\tablenotemark{a}.
}
\label{tab:clouds}
\tablenotetext{a}{
(1) name of the parent,
(2) area above the threshold [arcsec$^2$],
(3) distance from the most distant part of cloud to the center 
of the parent [kpc],
(4) position angle of the farthest part relative to the center
of the parent, north=0 and east=90 [degree]
(5) mean surface brightness of the clouds in R-band [mag arcsec$^{-2}$] 
including background continuum,
(6) peak \Ha\ surface brightness [mag arcsec$^{-2}$],
(7) total \Ha\ (isophotal) magnitude [mag],
(8) estimated total \Ha\ flux [erg s$^{-1}$ cm$^{-2}$],
(9) apparent morphological type of the clouds (see text)
}
\end{table}

\begin{table}
\begin{tabular}{|c|c|c|c|c|c|c|c|}
\hline
Name&
$r-i$&
$i$&
log(M/L)$_{\rm i}$&
M$_{\rm Bell}$&
M$_{\rm kcorrect}$&
M$_{\rm MPA}$&
type\\
(1) & (2) & (3) & (4) & (5) & (6) & (7) & (8)\\
\hline
GMP2374 &0.43 &12.6 &0.34 & 1300& 630\tablenotemark{b} & 1500 &1\\
GMP2559 &0.32 &14.8 &0.21 &  130&  43 &  100 &1\\
GMP2910 &0.23 &15.2 &0.11 &   73&  12 &   21 &2\\
GMP2923 &0.16 &16.7 &0.04 &   16&   4 &    8 &3\\
GMP3016 &0.18 &17.6 &0.06 &    7&   1 &   -- &3\\
GMP3071 &0.18 &16.0 &0.05 &   31&  11 &   19 &2\\
GMP3271 &0.17 &15.6 &0.05 &   44&   7 &    9 &1\\
GMP3779 &0.21 &14.2 &0.09 &  180&  33 &   47 &2\\
GMP3816 &0.27 &14.9 &0.16 &  110&  56 &  100 &1\\
GMP3896 &0.39 &13.8 &0.29 &  400&  93\tablenotemark{c} &  350 &2\\
GMP4017 &0.42 &13.4 &0.32 &  630& 300 &  730 &3(red)\\
GMP4060 &0.17 &16.6 &0.04 &   17&   4 &   -- &3\\
GMP4156 &0.39 &13.1 &0.29 &  770& 240 &  660 &3(red)\\
GMP4232 &0.13 &17.6 &0.01 &    6&   1 &   -- &3\\
\hline
\end{tabular}
\caption{Mass estimation of parent galaxies\tablenotemark{a}}
\label{tab:mass}
\tablenotetext{a}{
(1) name of the parent,
(2) $r-i$ color,
(3) $i$ magnitude,
(4) log$_{10}$ of i-band stellar mass-to-light ratio,
(5) estimated stellar mass [10$^8$ M$_{\sun}$] using \citet{Bell2003} formula,
(6) estimated stellar mass using kcorrect v4\_2 \citep{Blanton2007},
(7) MEDIAN stellar mass in MPA-JHU SDSS catalog,
and
(8) type.
}
\tablenotetext{b}{GALEX pipeline divided the galaxy into
multiple components. Their flux values were combined for the
total UV magnitude.}
\tablenotetext{c}{No FUV detection. We used FUV=25 with 
a large (9.99) error.}

\end{table}

\clearpage


\clearpage

{
\newcounter{stampfig}
\setcounter{stampfig}{\value{figure}}
\addtocounter{stampfig}{1}
\setcounter{figure}{0}
\renewcommand{\thefigure}{\thestampfig\Alph{figure}}

\begin{figure}[ht]
\includegraphics[scale=0.3,bb=0 0 633 581]{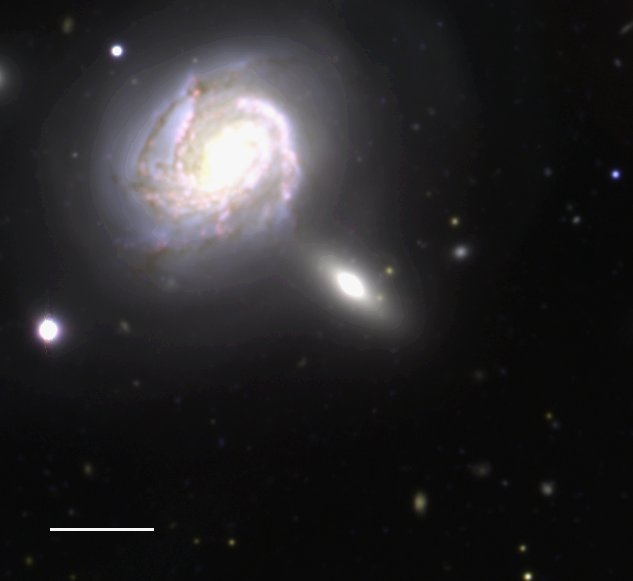}
\includegraphics[scale=0.3,bb=0 0 633 581]{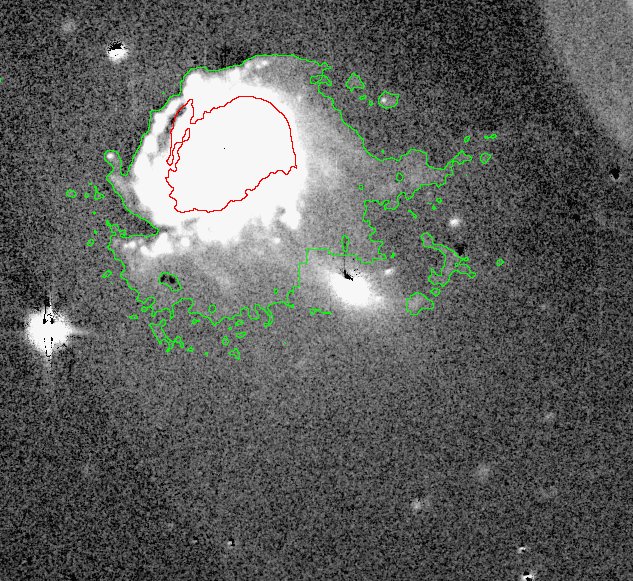}
\caption{
(left) B, R, and \NB\, three-color composite of GMP2374.
North is up and east to the left.
The relative B, R, and \NB\, scales are set so that 
typical Coma galaxies
without \Ha\ emission have gray color. The contrast is arbitrary.
The white bar represents a scale of 10 kpc (21 arcsec).
The pure red color represents \Ha\ emission without any stellar light,
the purple color represents \Ha\ emission with some stellar light,
and the blue color represents young blue stars without 
\Ha\ emission.
(right) \NB-R (\Ha) image of the same region as the left image.
The green contour represents the isophote of 
2.5$\times 10^{-18}$ erg s$^{-1}$ cm$^{-2}$ arcsec$^{-2}$.
The red contour represents the isophote of the R-band image
which corresponds to
the SDSS r-band $petroR50$ (the radius 
containing 50\% of the Petrosian flux).
A typical Coma galaxy without \Ha\ emission 
disappears in this image.
White indicates \NB\ flux excess and black indicates \NB\ flux 
deficiency.
As discussed in the text, the excess/deficiency correlated with 
the color of the underlying stellar components.
Redder and bluer underlying stellar components 
made \NB\ excess and \NB\ deficiency, respectively.
The contrast of \NB-R\ image (right) is the same 
for all the parents (Fig \ref{fig:postage2374}-\ref{fig:postage4232}).
}
\label{fig:postage2374}
\end{figure}

\begin{figure}[ht]

\includegraphics[scale=0.3,bb=0 0 569 1341]{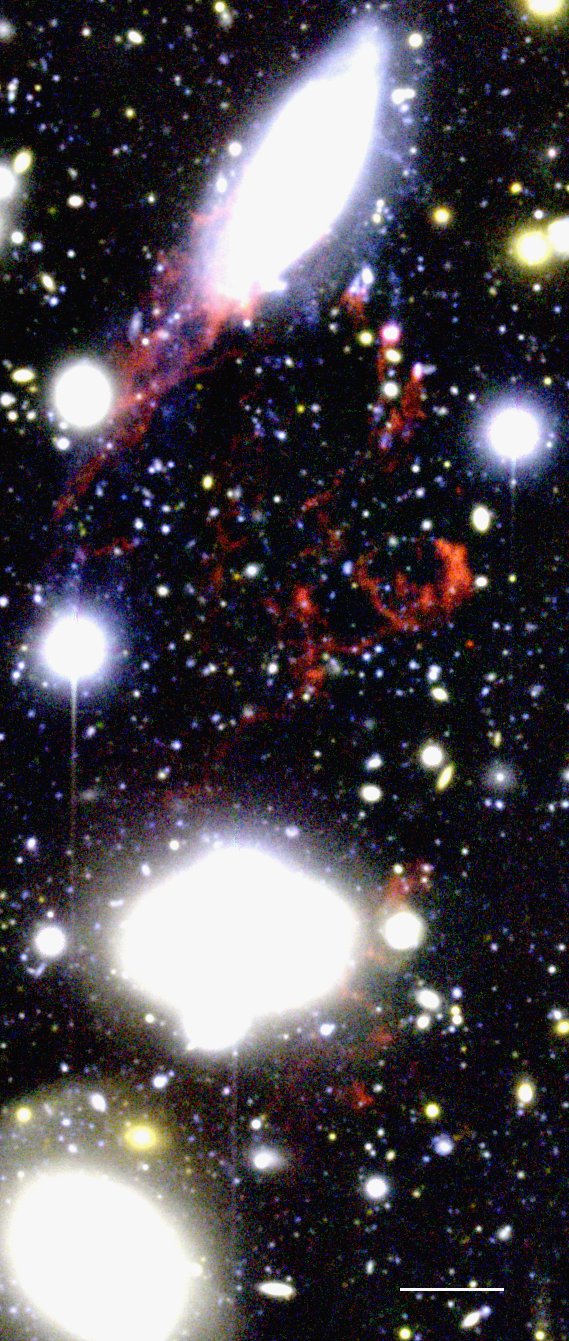}
\includegraphics[scale=0.3,bb=0 0 569 1341]{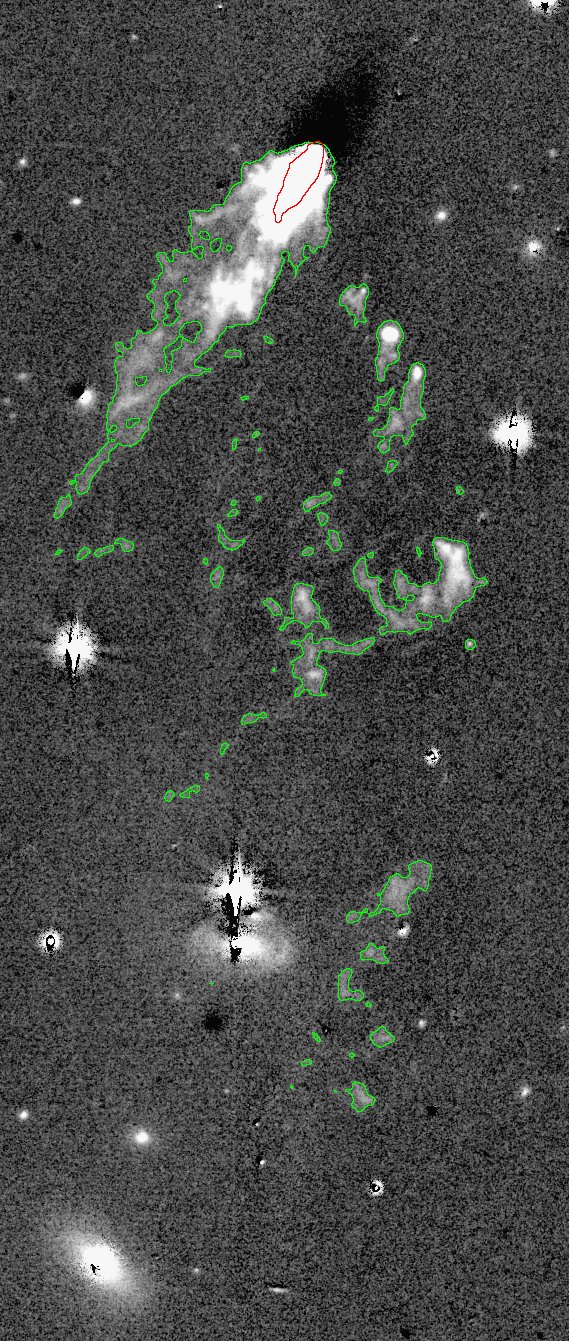}
\caption{
Same as Figure \ref{fig:postage2374}, but of GMP2559.
}
\label{fig:postage2559}
\end{figure}

\begin{figure}[ht]
\includegraphics[scale=0.3,bb=0 0 717 577]{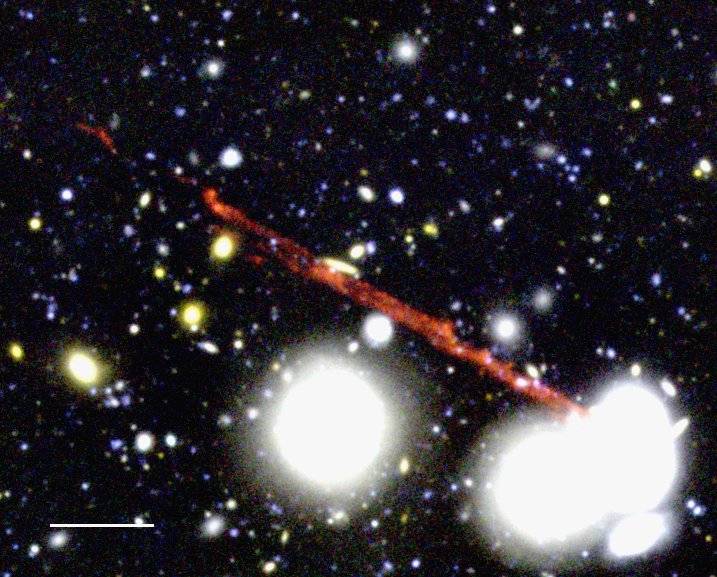}
\includegraphics[scale=0.3,bb=0 0 717 577]{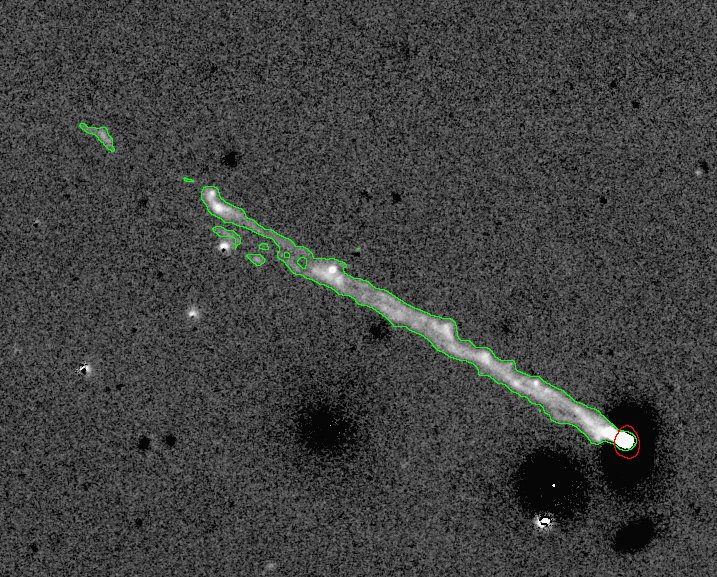}

\caption{
Same as Figure \ref{fig:postage2374}, but of GMP2910.
}
\label{fig:postage2910}
\end{figure}

\begin{figure}[ht]

\includegraphics[scale=0.3,bb=0 0 717 429]{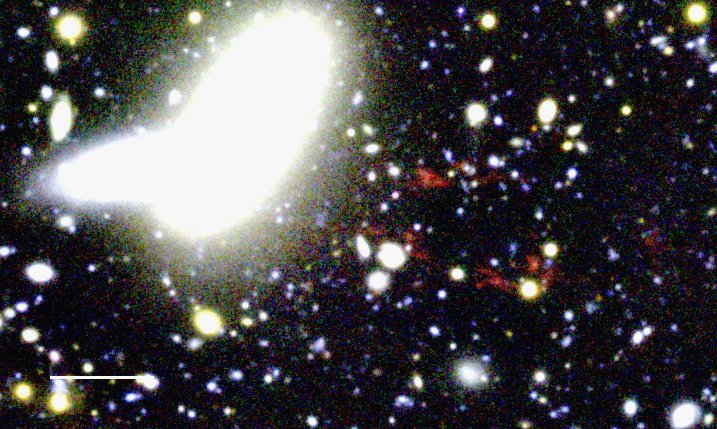}
\includegraphics[scale=0.3,bb=0 0 717 429]{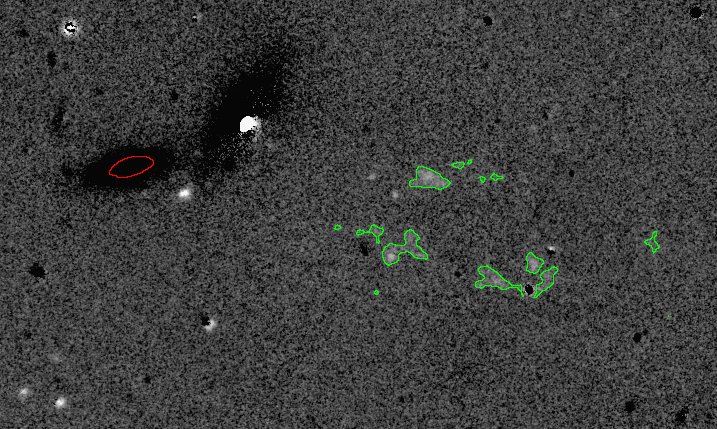}

\caption{
Same as Figure \ref{fig:postage2374}, but of GMP2923.
}
\label{fig:postage2923}
\end{figure}

\begin{figure}[ht]
\includegraphics[scale=0.3,bb=0 0 411 405]{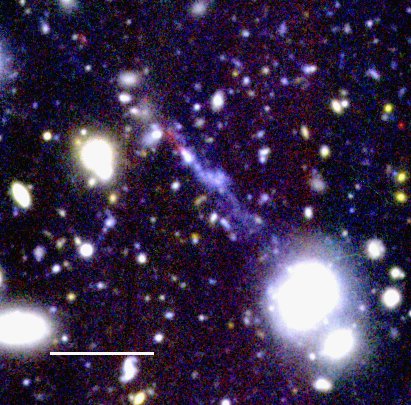}
\includegraphics[scale=0.3,bb=0 0 411 405]{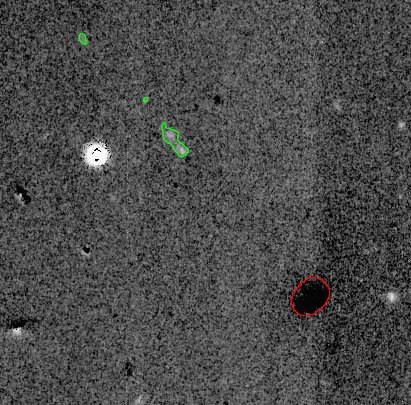}

\caption{
Same as Figure \ref{fig:postage2374}, but of GMP3016.
}
\label{fig:postage3016}
\end{figure}

\begin{figure}[ht]

\includegraphics[scale=0.3,bb=0 0 601 409]{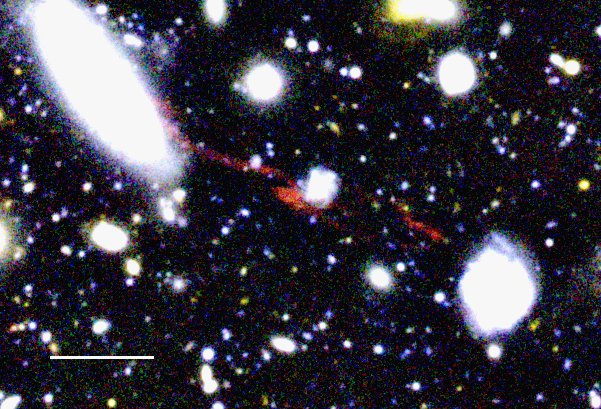}
\includegraphics[scale=0.3,bb=0 0 601 409]{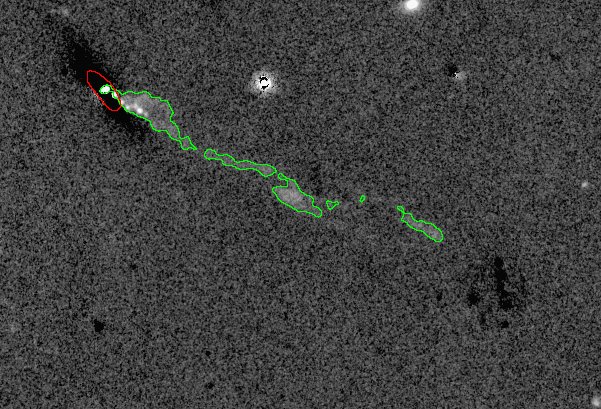}

\caption{
Same as Figure \ref{fig:postage2374}, but of GMP3071.
}
\label{fig:postage3071}
\end{figure}

\begin{figure}[ht]

\includegraphics[scale=0.3,bb=0 0 211 145]{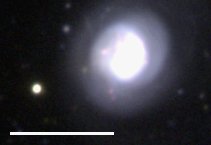}
\includegraphics[scale=0.3,bb=0 0 211 145]{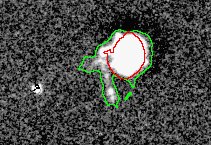}

\caption{
Same as Figure \ref{fig:postage2374}, but of GMP3271.
}
\label{fig:postage3271}
\end{figure}

\begin{figure}[ht]

\includegraphics[scale=0.3,bb=0 0 891 475]{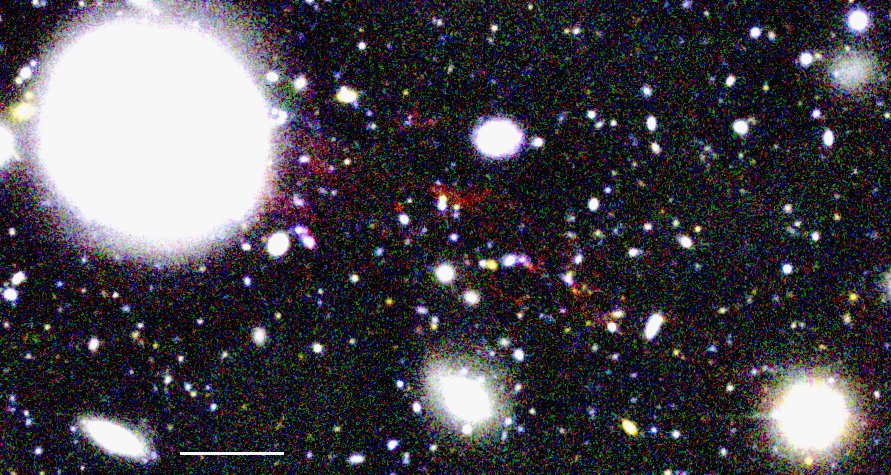}
\includegraphics[scale=0.3,bb=0 0 891 475]{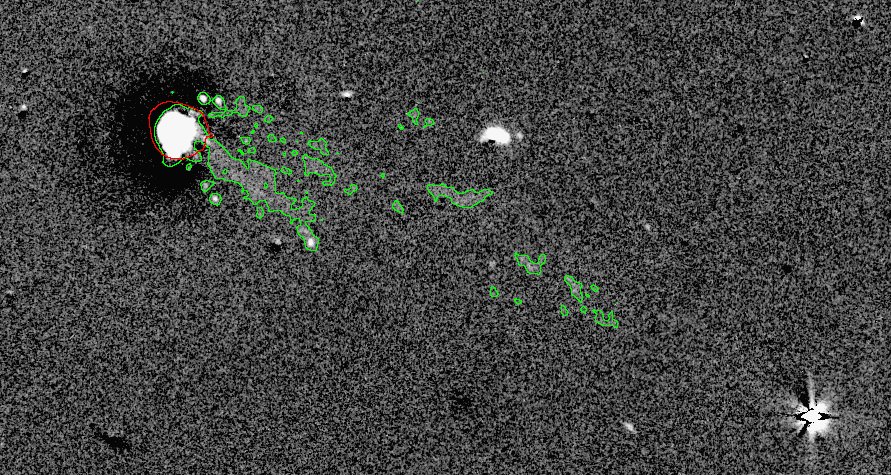}

\caption{
Same as Figure \ref{fig:postage2374}, but of GMP3779.
}
\label{fig:postage3779}
\end{figure}

\begin{figure}[ht]

\includegraphics[scale=0.3,bb=0 0 377 277]{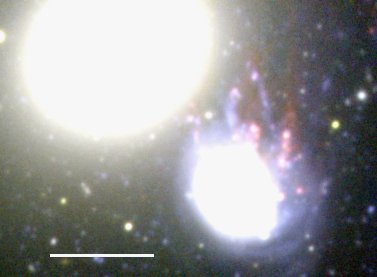}
\includegraphics[scale=0.3,bb=0 0 377 277]{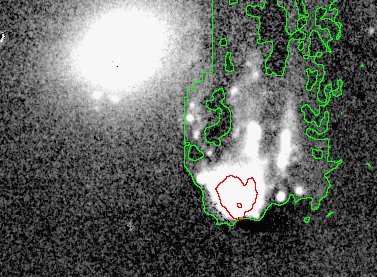}

\caption{
Same as Figure \ref{fig:postage2374}, but of GMP3816.
}
\label{fig:postage3816}
\end{figure}

\begin{figure}[ht]

\includegraphics[scale=0.3,bb=0 0 757 549]{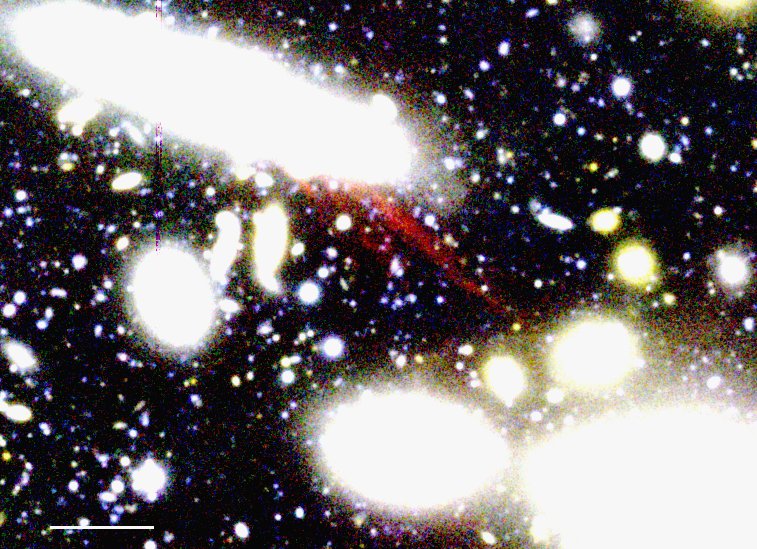}
\includegraphics[scale=0.3,bb=0 0 757 549]{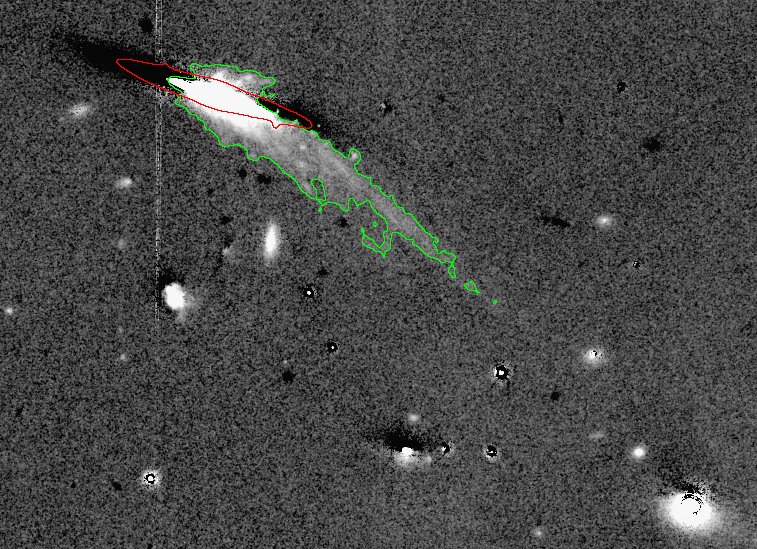}

\caption{
Same as Figure \ref{fig:postage2374}, but of GMP3896.
}
\label{fig:postage3896}
\end{figure}

\begin{figure}[ht]

\includegraphics[scale=0.3,bb=0 0 743 595]{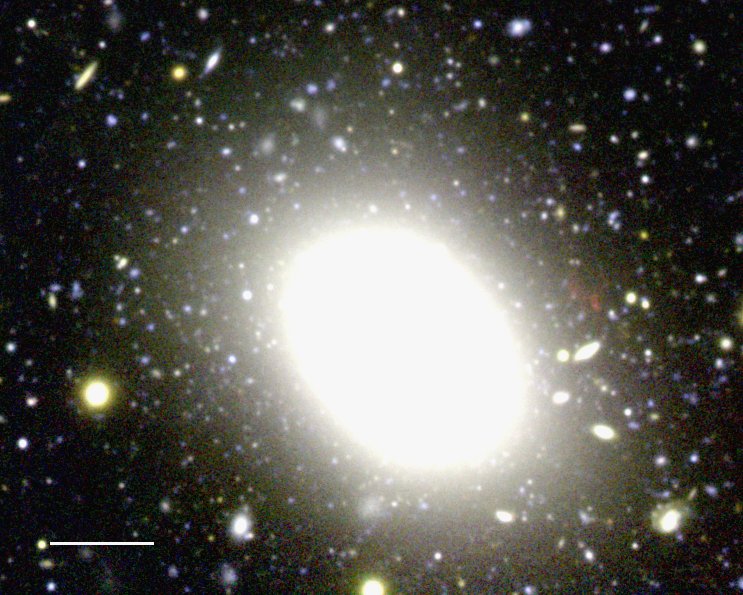}
\includegraphics[scale=0.3,bb=0 0 743 595]{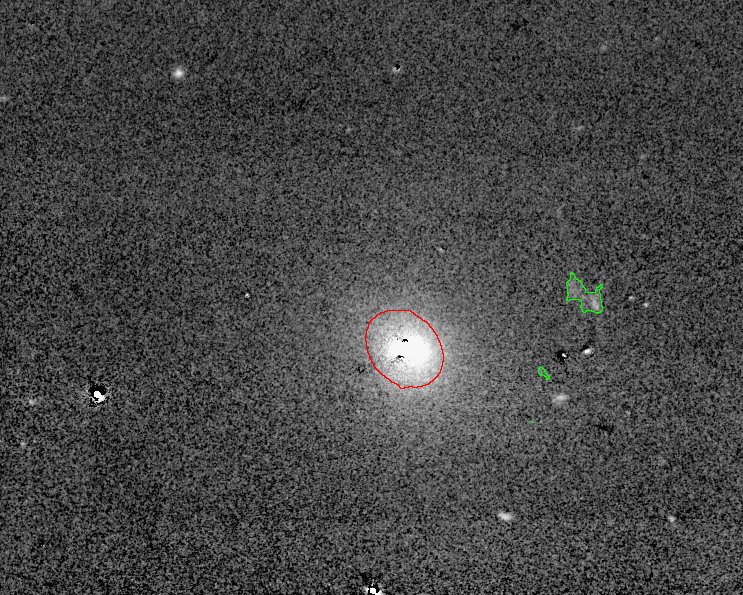}

\caption{
Same as Figure \ref{fig:postage2374}, but of GMP4017.
}
\label{fig:postage4017}
\end{figure}

\begin{figure}[ht]

\includegraphics[scale=0.3,bb=0 0 531 1083]{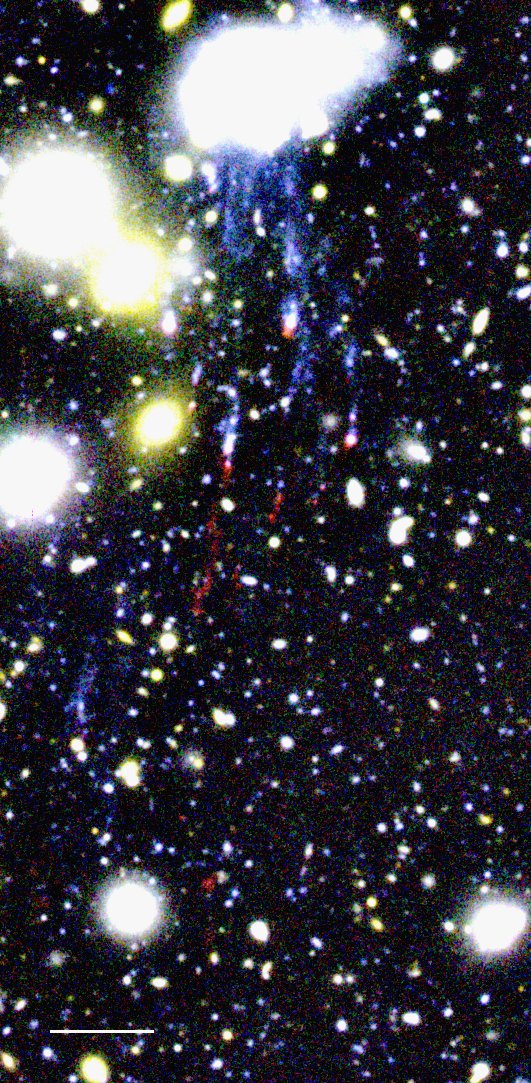}
\includegraphics[scale=0.3,bb=0 0 531 1083]{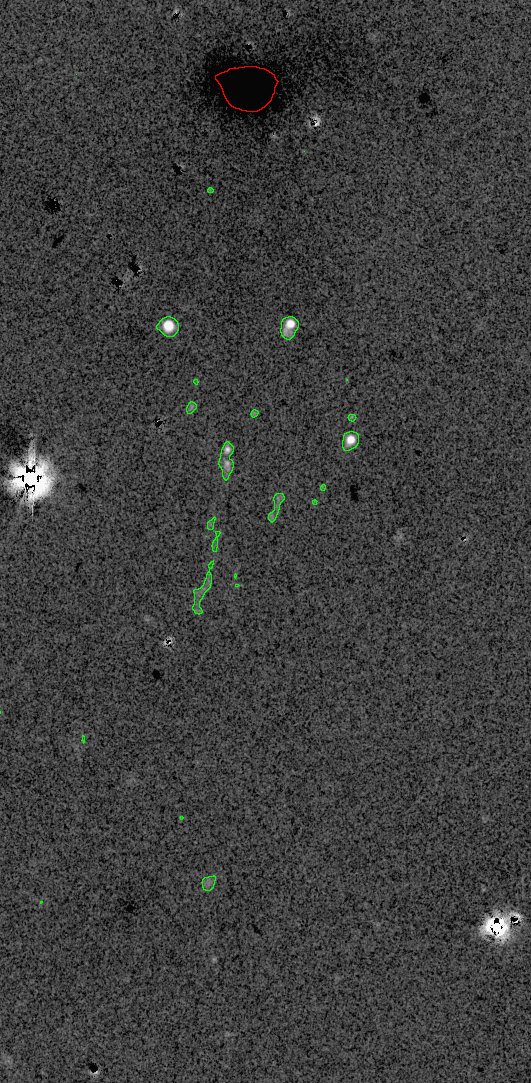}

\caption{
Same as Figure \ref{fig:postage2374}, but of GMP4060.
}
\label{fig:postage4060}
\end{figure}

\begin{figure}[ht]

\includegraphics[scale=0.3,bb=0 0 457 459]{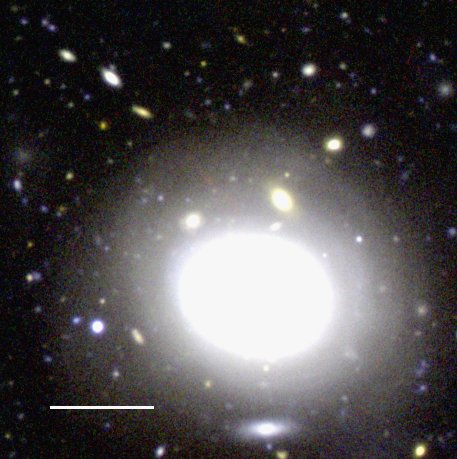}
\includegraphics[scale=0.3,bb=0 0 457 459]{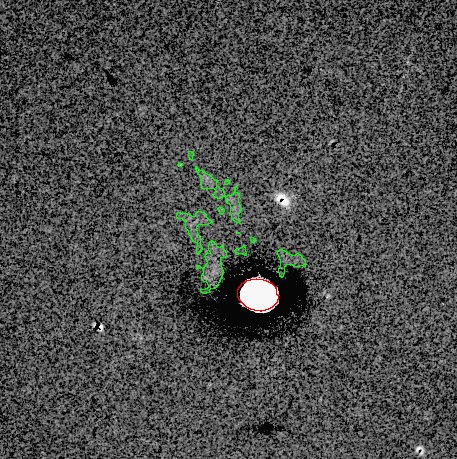}

\caption{
Same as Figure \ref{fig:postage2374}, but of GMP4156.
}
\label{fig:postage4156}
\end{figure}

\begin{figure}[ht]

\includegraphics[scale=0.3,bb=0 0 249 291]{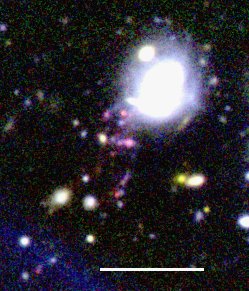}
\includegraphics[scale=0.3,bb=0 0 249 291]{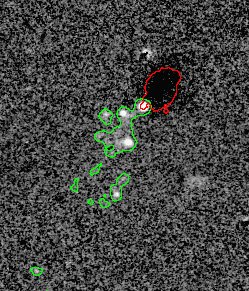}

\caption{
Same as Figure \ref{fig:postage2374}, but of GMP4232.
}
\label{fig:postage4232}
\end{figure}

\clearpage
\setcounter{figure}{\value{stampfig}}
}

\clearpage

\begin{figure}
\includegraphics[scale=0.3,bb=14 14 968 1940]{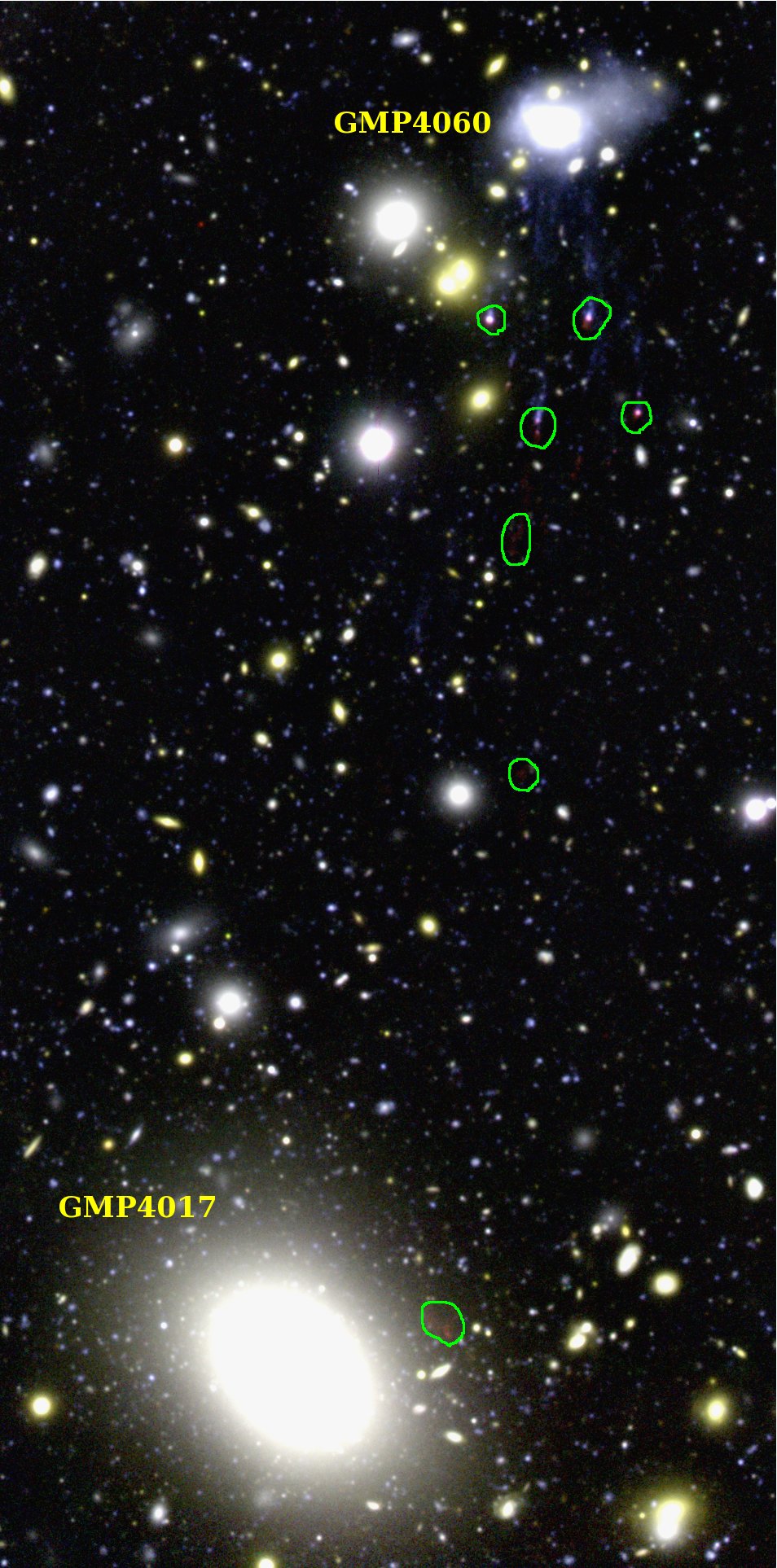}
\caption{
Three-color composite between GMP4060 and GMP4017.
The contrast is the same as in Figure \ref{fig:postage4017} left.
and the stellar streams seen in Figure \ref{fig:postage4060} 
are difficult to recognize.
Noticeable \Ha\ clouds are marked by green open shapes.
}
\label{fig:GMP4017a}
\end{figure}

\clearpage

\begin{figure}[h]
\includegraphics[scale=0.7,bb=20 17 592 779]{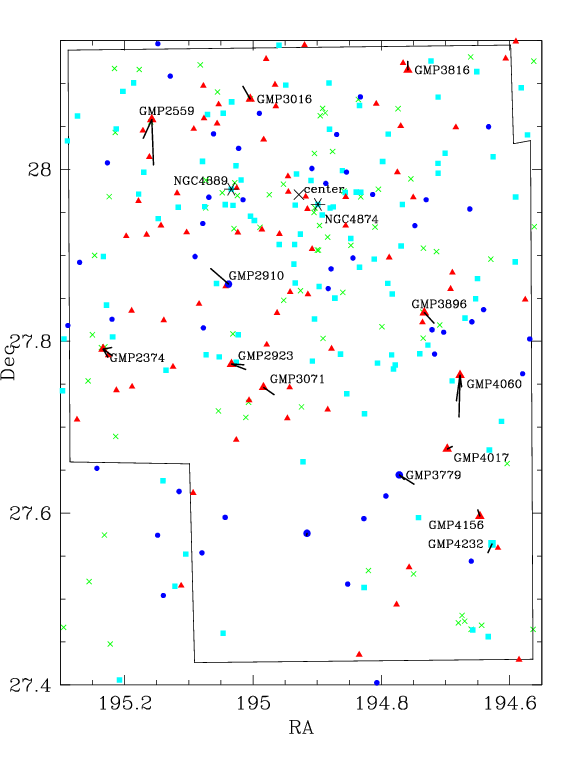}
\caption{
The position angle and extension of \Ha\ clouds in the 
observed region. The cluster center is shown as a cross
and cD galaxies (NGC4874 and NGC4889) are shown by the two asterisks.
The redshift by SDSS and by Hectospec for the two parents 
are represented as 
blue circles(0.015$<$z$<$0.02), cyan squares(0.02$\leq$z$<$0.025), 
red triangles(0.025$\leq$z$<$0.035), and green crosses(no data), respectively.
The parent galaxies are shown as large symbols with 
solid lines, indicating the direction and distance to
the tips of the clouds. 
The length of the lines is not zoomed and the scale on the 
margin applies.
}
\label{fig:FC2}
\end{figure}

\begin{figure}[ht]
\includegraphics[angle=-90,scale=0.3,bb=22 16 596 784]{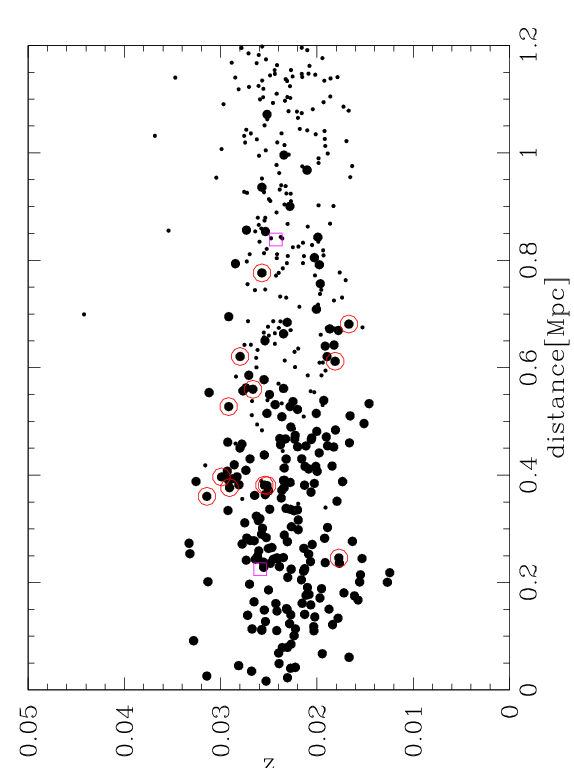}
\includegraphics[angle=-90,scale=0.3,bb=22 16 596 784]{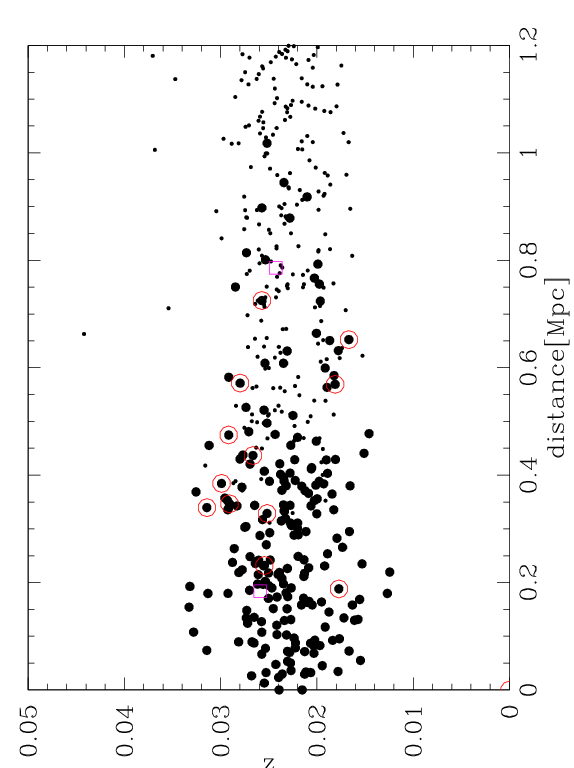}
\caption{
(left) 
Redshift vs the distance from the cluster center.
All galaxies brighter than $r=17.8$ mag and with SDSS spectroscopic data 
within 42.2 arcmin (which corresponds to 1.2 Mpc at the Coma distance)
from the Coma cluster center are plotted as small dots.
The galaxies in our observed regions are shown as filled circles.
The 12 parent galaxies with SDSS spectra are marked with red
open circles, and the two parent galaxies with Hectospec spectra are
marked with magenta open squares.
(right) Same as left, except the distance from the 
nearest cD is on the abscissa.
}
\label{fig:d_z}
\end{figure}

\begin{figure}[ht]
\includegraphics[angle=-90,scale=0.55,bb=22 16 596 784]{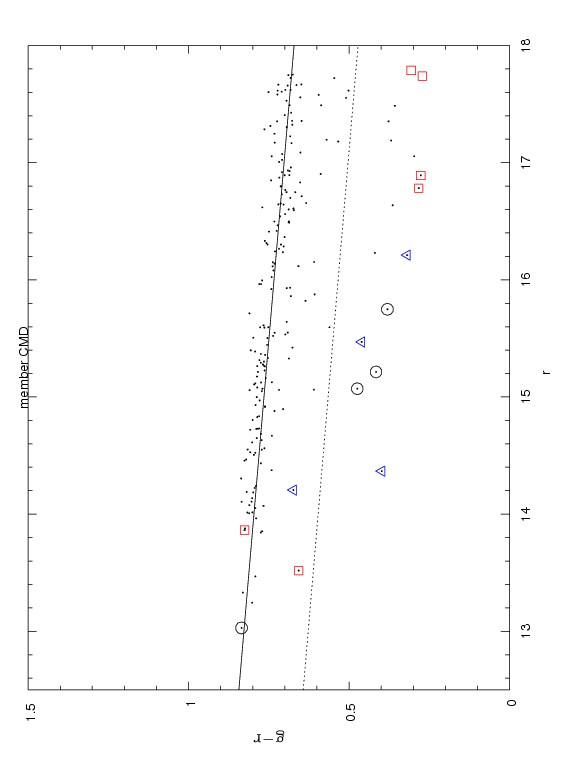}
\caption{
Color magnitude diagram of member galaxies 
brighter than $r=17.8$ mag in SDSS.
The large symbols show the parent galaxies of the \Ha\ clouds.
The squares without a dot represent GMP3016 and GMP4232,
for which spectroscopic data are unavailable in SDSS.
The colors and symbols represent the morphological type of 
the parent-cloud connection; 
black open circles(\#1; connected \Ha\ clouds with disk star formation), 
blue open triangles(\#2; connected \Ha\ clouds without disk star formation), 
and red squares(\#3; detached \Ha\ clouds).
The solid line is the best-fit of the color-magnitude relation (CMR) 
of passive galaxies. The broken line is the 0.2 mag shifted CMR,
which is a demarcation line between blue and red galaxies.
}
\label{fig:colmag}
\end{figure}

\clearpage

\appendix
{
\setcounter{figure}{0}
\setcounter{table}{0}

\section{Suprime-Cam flux calibration using SDSS}
\label{sec:photocal}

\renewcommand{\thefigure}{A\arabic{figure}}
\renewcommand{\thetable}{A\arabic{table}}

We took a special care for the flux calibration.
The B-band of Suprime-Cam has different transmission 
from the standard Johnson-B, because of the low throughput of
CCDs at shorter wavelengths.
In addition, the R-band of Suprime-Cam has
different transmission from Kron-Cousins-R
as it has a much shallower range
\footnote{
http://www.naoj.org/Observing/Instruments/SCam/sensitivity.html}.

We adopted to use the SDSS DR7 catalogs\citep{DR7}, 
which cover the whole of our observation area.
As many stars in the observed field had photometric data
in SDSS DR7, we calibrated the zero-point
of our data to archieve the best-match.
We retrieved type=6(stellar objects) of $r<21$ objects
for the calibration.
The colors of typical stars were calculated 
using the Bruzual-Persson-Gunn-Stryker Atlas
\footnote{ftp://ftp.stsci.edu/cdbs/grid/bpgs/}.
Using the filter transmission of SDSS DR7 and 
that of the Suprime-Cam filters, we plotted $g-B$(Suprime) vs. 
$g-r$, $r-R$(Suprime) vs. $g-r$, and $i-i$(Suprime) vs. $r-i$.
Then, we fitted the loci of stars with a second order
polynomial as
\[
SDSS-Suprime = c_0 + c_1 (color) + c_2 (color)^2.
\]
The coefficients are shown in Table \ref{tab:colconv_coeff}, and 
the fit is also shown as 
Figures \ref{fig:coldepold} and \ref{fig:coldepnew}.

Replacement of the Suprime-Cam CCDs in July 2008 
\citep{Kamata2008} affected the color terms.
The systematic differences between the old and new CCDs are
negligible ($<0.01$ mag) for R and i band, but
$\sim 0.05$ mag for B-band (Table \ref{tab:coldiff}). 
This was because of the change in quantum efficiency at 
$\lambda<4500$\AA.
Therefore we made two B-band images;
one from old CCD data and the other from new CCD data.
As the latter covers a part of the field, the three color
images in this paper were created with the old CCD B-band data.

The cross-identification of stars in each FITS dataset was
performed using celestial coordinates. 
We used wcstools\citep{Mink2002} and 
the USNO-B1.0 catalog \citep{Monet2003}
to calibrate the world coordinate system (WCS).
The flux measurement was performed with SExtractor \citep{Bertin1996}.
Removing saturated or blended objects, 
we estimated the shift between the observed data and the model locus 
using the mode of the difference.

The AB magnitude of the \NB-band was also calibrated with SDSS
in the same manner. 
We assumed that most of the stars did not have strong 
features around 6710\AA$\pm$ 100\AA,
and hence that the $r-\NB$(Suprime) vs. $g-r$ plot would work.
We also used spectroscopic data of stars in our observed field
to calculate $r-\NB$(Suprime) color.
The values were consistent with the model locus of BPGS stars.

This method is applicable to any Suprime-Cam data
where SDSS has a photometric catalog.
However, this method has the drawback that
Galactic extinction may add systematic error,
because the model locus is free from the extinction.
In the Coma region, however, the Galactic extinction for
extragalactic objects is $A_B=0.035$, from the model of 
\citet{Schlegel1998} in NED. 
The extinction for Galactic stars must be 
much less than this value, and therefore we 
neglected the effect in this study.

\begin{table}
\begin{tabular}{|c|c|c|c|c|c|}
\hline
SDSS-Suprime& SDSS color& range &$c_0$&$c_1$&$c_2$\\

\hline
$g-B$(new)  &$g-r$&-0.6$<g-r<$1.4&-0.0232335&-0.174584&-0.0138875\\
$g-B$(old)  &$g-r$&-0.6$<g-r<$1.4&-0.0317803&-0.212379&-0.0178932\\
\hline
$r-R$(new)  &$g-r$&-0.6$<g-r<$1.4&-0.0188297&0.149276&-0.0128141\\
$r-R$(old)  &$g-r$&-0.6$<g-r<$1.4&-0.0191869&0.151382&-0.012914\\
\hline
$i-i$(new)  &$r-i$&-0.4$<r-i<$2.5&-0.00488511&0.101026&0.00341974\\
$i-i$(old)  &$r-i$&-0.4$<r-i<$2.5&-0.00472411&0.102785&0.00326376\\
\hline
$r-\NB$(new)&$g-r$&-0.6$<g-r<$1.6&-0.0214062&0.250467&-0.0361111\\
$r-\NB$(old)&$g-r$&-0.6$<g-r<$1.6&-0.021407&0.250514&-0.0361369\\
\hline
\end{tabular}
\caption{The coefficients of best-fit color conversion polynomials}
\label{tab:colconv_coeff}
\end{table}

\begin{table}
\begin{tabular}{|c|c|}
\hline
SDSS-Suprime& difference\\
\hline
$g-B$& 0.12 \\
$r-R$& 0.01 \\
$i-i$& 0.02 \\
$r-\NB$& 0.00 \\
\hline
\end{tabular}
\caption{The maximum color difference of BPGS stars in 
old and new CCD}
\label{tab:coldiff}
\end{table}

\begin{figure}[ht]
\includegraphics[angle=-90,scale=0.3,bb=22 16 596 784]{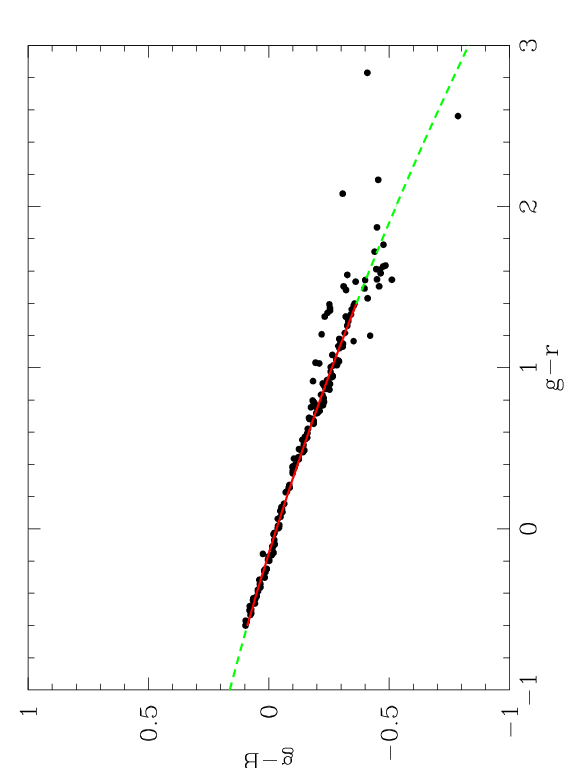}
\includegraphics[angle=-90,scale=0.3,bb=22 16 596 784]{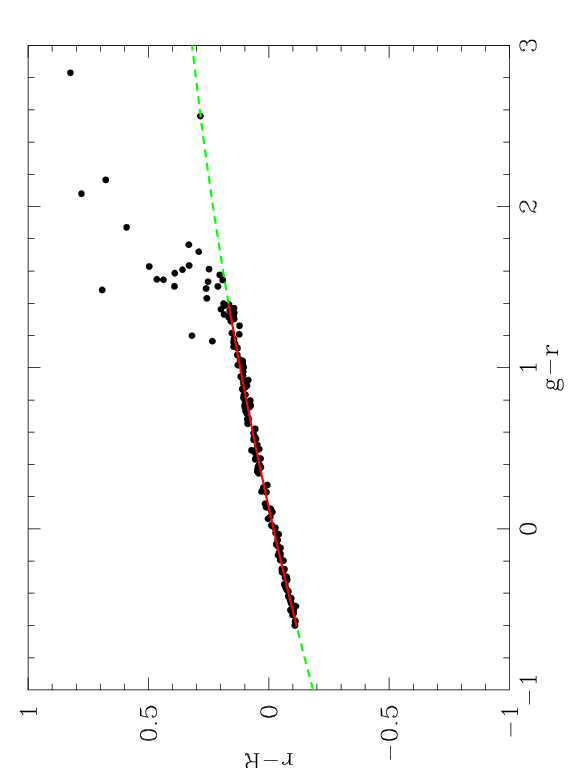}

\includegraphics[angle=-90,scale=0.3,bb=22 16 596 784]{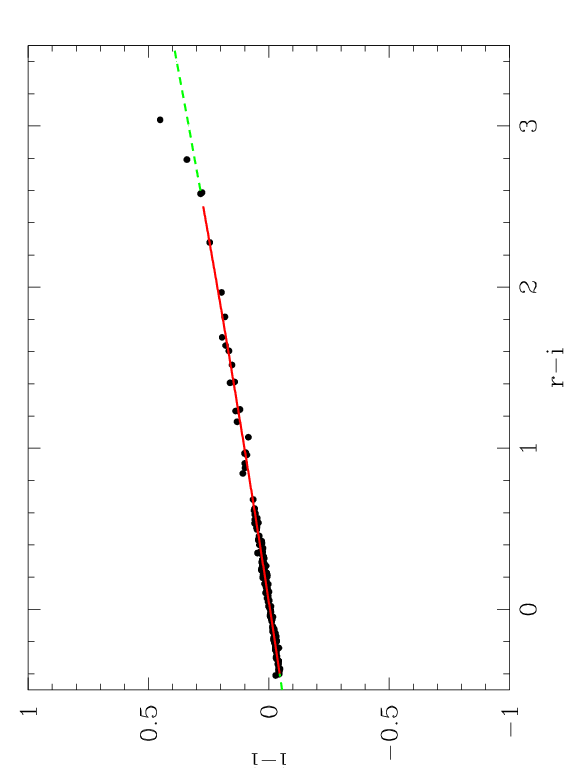}
\includegraphics[angle=-90,scale=0.3,bb=22 16 596 784]{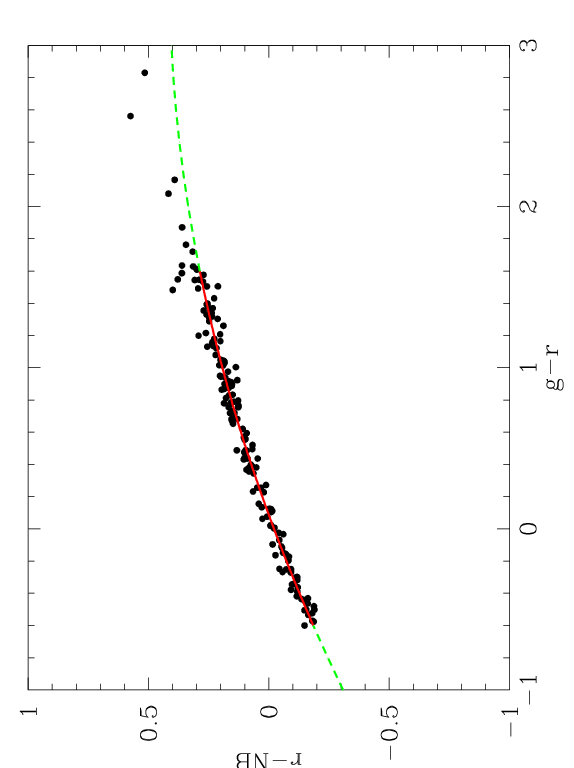}
\caption{
SDSS color vs. SDSS-Suprime color of BPGS stars for
old Suprime-Cam CCDs.
The line represents the regression curve of the relation,
where the range used in the calibration is shown as a red solid line
and the other part is shown as a green broken line.
}
\label{fig:coldepold}
\end{figure}

\begin{figure}[ht]
\includegraphics[angle=-90,scale=0.3,bb=22 16 596 784]{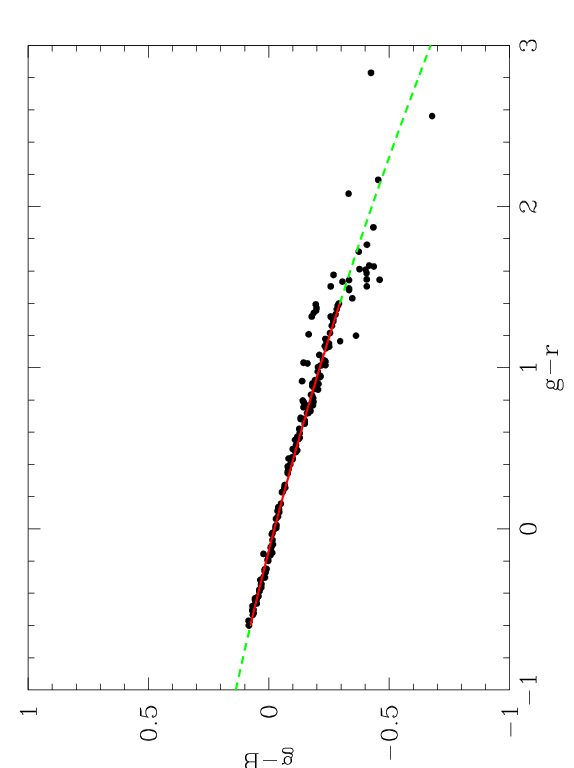}
\includegraphics[angle=-90,scale=0.3,bb=22 16 596 784]{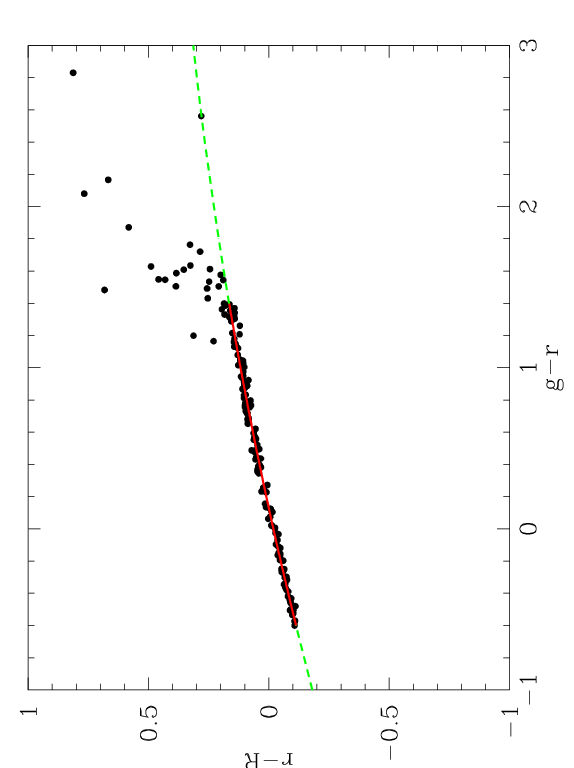}

\includegraphics[angle=-90,scale=0.3,bb=22 16 596 784]{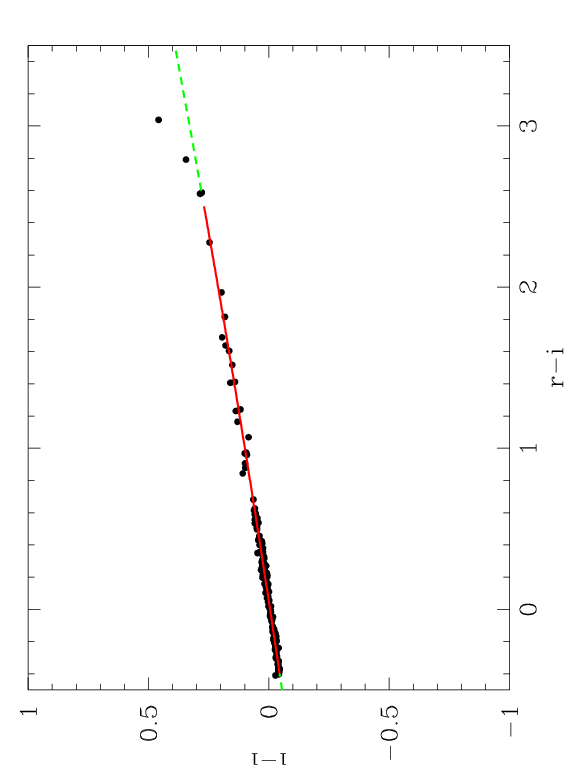}
\includegraphics[angle=-90,scale=0.3,bb=22 16 596 784]{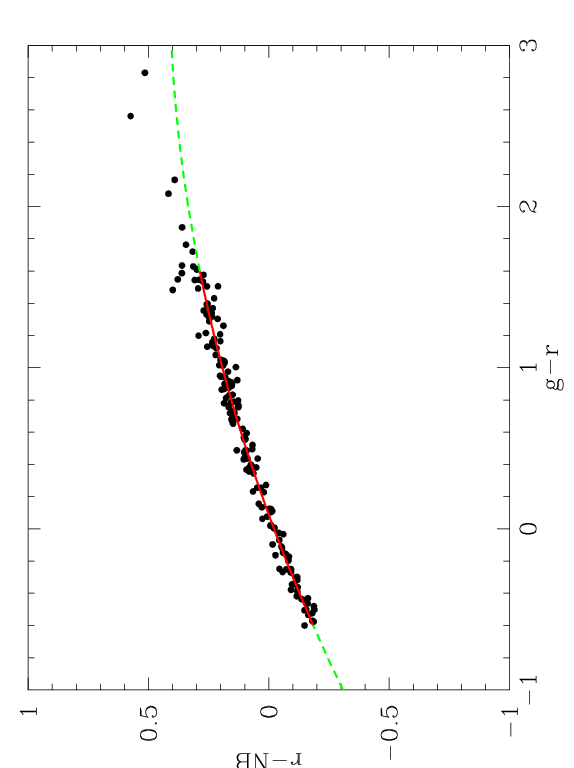}
\caption{
Same as Figure \ref{fig:coldepold} for new Suprime-Cam CCDs.
}
\label{fig:coldepnew}
\end{figure}
}
\clearpage

\end{document}